\documentclass[12pt, referee]{aastex}
\usepackage[dvips]{color}

\newcommand{\gr}{\textcolor[rgb]{0,0,0}}

\def\gtsima{$\;\buildrel > \over \sim \;$}
\def\simgt{\lower.5ex \hbox{\gtsima}}
\def\ltsima{$\;\buildrel < \over \sim \;$}
\def\simlt{\lower.5ex \hbox{\ltsima}}

\begin{document}

\title{Detection of vibrational emissions from the helium hydride ion (HeH$^+$) in the planetary nebula NGC 7027}

\author{David A. Neufeld\altaffilmark{1}, Miwa Goto\altaffilmark{2}, \gr{T.\ R.}\ Geballe\altaffilmark{3}, Rolf G\"usten\altaffilmark{4}, Karl M. Menten\altaffilmark{4}, and
Helmut Wiesemeyer\altaffilmark{4}}

\altaffiltext{1}{Department of Physics \& Astronomy, Johns Hopkins Univ., Baltimore, MD 21218, USA}
\altaffiltext{2}{Universit\"ats-Sternwarte M\"unchen, Ludwig-Maximilians-Universit\"at, 
D-81679 M\"unchen, Germany}
\altaffiltext{3}{Gemini Observatory, 670 N.\ A'ohoku Place, Hilo, HI 96720, USA}
\altaffiltext{4}{Max-Planck-Institut f\"ur Radioastronomie, Auf dem H\"ugel 69, 53121 Bonn, Germany}

\begin{abstract}

We report the detection of emission in the v$=1-0$ $P(1)$ (3.51629$\,\mu$m) and $P(2)$ (3.60776$\,\mu$m) rovibrational 
lines of the helium hydride cation (HeH$^+$) from the planetary nebula NGC 7027.
These detections were obtained with the iSHELL spectrograph on NASA's Infrared Telescope Facility (IRTF) on Maunakea.  They confirm the discovery of HeH$^+$ reported recently by G\"usten et al.\ (2019), who 
used the GREAT instrument on the SOFIA airborne observatory
to observe its pure rotational $J=1-0$ transition at 149.137$\,\mu$m.
The flux measured for the HeH$^+$ v$=1-0$ $P(1)$ line is in good agreement with 
our model for the formation, destruction and excitation of HeH$^+$ in NGC 7027.  
The measured strength of the $J=1-0$ pure rotational line, however, exceeds the model prediction significantly, 
as does that of the v$=1-0$ $P(2)$ line, by factors of 2.9 and 2.3 respectively.
Possible causes of these discrepancies are discussed.
Our observations of NGC 7027, covering the 3.26 -- 3.93$\,\mu$m spectral region, 
have led to the detection of more than sixty spectral lines including nine rovibrational emissions from CH$^+$.  
The latter are detected for the first time in an astronomical source. 

\end{abstract}


\section{Introduction}

The first astrophysical detection of the helium hydride ion (HeH$^+$) was reported recently by G\"usten et al.\ (2019; hereafter G19), who used the German Receiver for Astronomy at Terahertz Frequencies (GREAT) on the Stratospheric Observatory for Infrared Astronomy (SOFIA) to detect its fundamental rotational transition at 2010.184 THz (149.137$\mu$m) toward the planetary nebula NGC 7027.  This came more than four decades after HeH$^+$ was first recognized as potentially-detectable interstellar molecule (Dabrowski \& Herzberg 1978; Black 1978), and followed multiple unsuccessful searches at both infrared and far-infrared wavelengths (Moorhead et al.\ 1988; Dinerstein \& Geballe 2001; Liu et al.\ 1997)

The HeH$^+$ molecular ion had been discovered in the laboratory in 1925 (Hogness \& Lunn 1925), when mass spectroscopy revealed the presence of an ion of charge-to-mass ratio $q/m = \displaystyle{\scriptsize 1 \over 5}(e/m_p)$ produced by a discharge in a mixture of hydrogen and helium.  
HeH$^+$ is isoelectronic with H$_2$, {with a $^1 \Sigma$ ground state}.   Indeed, H$_2$ and HeH$^+$ constitute one example of several isoelectronic pairs or multiplets known in astrochemistry where exotic ionized species  have been discovered that have a familiar neutral counterpart: other examples include CF$^+$ (Neufeld et al.\ 2006) and CN$^-$ (Agundez et al.\ 2010), both isoelectronic with CO; as well as ArH$^+$ (Barlow et al.\ 2013; Schilke et al.\ 2014), $\rm H_2Cl^+$ (Lis et al. 2010) and $\rm HCO^+$ (Klemperer 1970), molecular ions that are isoelectronic with HCl, H$_2$S and HCN respectively. 

\gr{HeH$^+$ was first mentioned as a potential interstellar molecule by Wildt (1949).   Dabrowski \& Herzberg (1976) subsequently suggested} that HeH$^+$ vibrational emissions might be responsible for spectral features that had been detected (Merrill et al.\ 1975) in low resolution infrared spectra of the young planetary nebula \gr{(PN)} NGC 7027.  While those spectral \gr{features} were in fact associated with polycyclic aromatic hydrocarbons, the suggestion prompted several theoretical studies that investigated the formation of HeH$^+$ in planetary nebulae Black 1978; Flower \& Roeuff 1979; Roberge \& Dalgarno 1982; Cecchi-Pestellini \& Dalgarno 1993). \gr{These} concluded that potentially-detectable HeH$^+$ abundances might be formed by the radiative association of 
He$^+$ with H in the transition regions of planetary nebulae (i.e. between the fully-ionized and neutral zones).
In \gr{such environments} surrounding a hot UV-emitting central star, there can be a small overlap between the region containing ionized helium and that containing neutral hydrogen.  This (somewhat counterintuitive) possibility arises because the photoelectric absorption cross-section of hydrogen falls rapidly with energy above the ionization threshold (13.6 eV), so that ultraviolet photons with energies slightly greater than the ionization potential of He (24.6 eV) penetrate more deeply into the neutral zone than those with energies just above the ionization potential of H (13.6~eV); the helium Str\"omgren sphere therefore is slightly larger than the hydrogen Str\"omgren sphere, despite (in fact, because of) the higher ionization potential of He.
  
In addition to these theoretical studies of HeH$^+$ in planetary nebulae ({PNe}), HeH$^+$ has also been included in models for the formation of molecules in the primordial Universe, prior to the formation of stars and before the advent of stellar nucleosynthesis (e.g. Stancil et al.\ 1998).  Current models (e.g.\ Galli \& Palla 2013; their Figure 3c) predict HeH$^+$ to be the very first molecule to form (followed rapidly by $\rm He_2^+$) once helium atoms start \gr{to} recombine at redshift $z \sim 7000$.  In the early Universe, HeH$^+$ is formed primarily by the radiative association of He with H$^+$, which is much slower than the reaction of He$^+$ with H that dominates in planetary nebula.  Here, the overlap is between He and H$^+$; with its higher ionization potential, helium recombines before hydrogen in the cooling Universe. 
HeH$^+$ plays an important role in the subsequent molecular evolution of the Universe, since it is destroyed primarily by proton transfer to atomic hydrogen.  This reaction forms H$_2^+$, which can subsequently undergo charge transfer with H to form H$_2$.
  
\gr{The theoretical studies addressing HeH$^+$ in PNe} motivated several observational searches for HeH$^+$ emission from NGC 7027, a young PN that was recognized as a particularly promising target because of its very hot central star and the high density of the surrounding nebula.  Upper limits on its emission were obtained from ground-based observatories operating in the infrared L-band by Moorhead et al. (1988), who searched for the $v = 1-0\,R(0)$ {rovibrational} line at 3.364~$\mu$m, and by Dinerstein \& Geballe (2001), who targeted the (stronger) $v = 1-0\,P(2)$ line at 3.608~$\mu$m.  

At far-infrared wavelengths, observations of NGC 7027 performed with the {\it Infrared Space Observatory} ({\it ISO}) revealed (Liu et al.\ 1997) an emission feature at $149.18 \pm 0.06$~$\mu$m.  Given the relatively-poor spectral resolution ($\Delta \lambda = 0.6 \,\mu$m) of these observations, which were obtained with the Long Wavelength Spectrometer (LWS) in its grating mode, the wavelength of the observed emission feature was consistent with either the $J=1-0$ transition of HeH$^+$ (149.14$\mu$m) or the $^2\Pi_{3/2}\,J=3/2 \rightarrow 1/2$ $\lambda$ doublet transitions (149.09 and $149.39\,\mu$m) of CH.  Indeed, the strength of the observed $149.18\,\mu$m emission feature was entirely consistent with that expected from CH, given the measured strength of another CH transition detected simultaneously at 180.7$\,\mu$m.  Accordingly, Liu et al.\ (1997) were only able to derive an upper limit on any contribution from HeH$^+$ $J=1-0$.  It was the GREAT spectrometer on SOFIA that was the first instrument with the sensitivity and spectral resolution to detect the HeH$^+$ fundamental rotational transition and discriminate it from the nearby CH $\lambda$ doublet (G19).

To follow up the recent discovery of astrophysical HeH$^+$, we have used the iSHELL spectrograph (Rayner et al.\ 2016) on NASA's Infrared Telescope Facility (IRTF) to conduct a sensitive search for HeH$^+$ vibrational emissions from NGC 7027.  In Sections 2 and 3 below we describe the observations we performed and the methods used to reduce the data.  In Section 4, we present the observational results, which provide a clear detection of HeH$^+$ vibrational emissions.  In Section 5, we compare our measurements of the HeH$^+$ line fluxes with the predictions of a model for the source.  A brief summary follows in Section 6.  In Appendix A, we present the full $3.28 - 3.93\,\mu$m spectrum obtained using iSHELL on IRTF.

\section{Observations}

\begin{figure}
\includegraphics[scale=0.8,angle=-0]{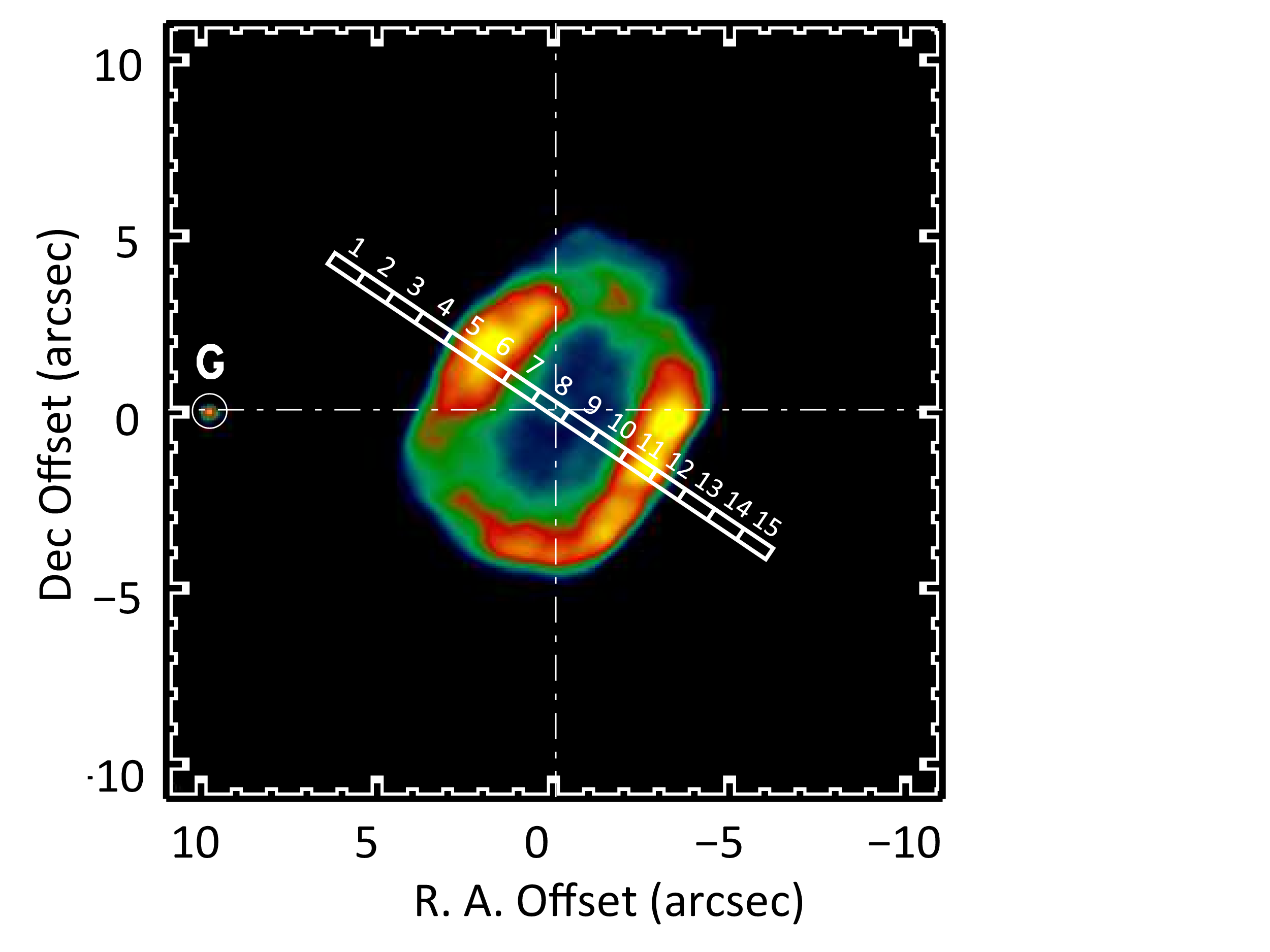}
\caption{$K$-band image of NGC 7027, obtained with the slit viewing camera.  White boxes indicate the aperture extraction regions adopted, with the numbering system used in the text.  \gr{The R.A.\ and Dec.\ offsets are given relative to the central star at $\alpha = \rm 21^h\,07^m\,1\fs 793, \delta=42\degr\,14^{\prime\prime}\,9\farcs 79$ (J2000).  ``G" indicates the star used for guiding.}}
\end{figure} 

The observations were performed at the IRTF on the nights of \gr{2019 July 13
and September 6 UT} using the iSHELL spectrograph (Rayner
et al. 2016).  The {\tt Lp1} and {\tt Lp2} grating settings were used to cover
\gr{the} 3.265 to 3.656\,$\mu$m and 3.580 to
3.933 \,$\mu$m spectral regions, respectively, \gr{on the first and
second nights of observation.}
The former \gr{setting} covers the HeH$^+$ $v=1-0$ $R$(1)
(3.3021\,$\mu$m)\footnote{Here, and elsewhere in the paper, all wavelengths are given in vacuo}, 
$R$(0) (3.3641\,$\mu$m), $P$(1)
(3.5163\,$\mu$m) and $P$(2) (3.6078\,$\mu$m) rovibrational lines, 
and the latter covers $v=1-0$
$P$(2), $P$(3) (3.7105\,$\mu$m), and $P$(4) (3.8255\,$\mu$m) (Bernath \& Amano 1982).
The \gr{0\farcs375 wide} slit was used on both nights to deliver
spectra with a spectral resolving power $\lambda/\Delta \lambda=80,000$, corresponding to a \gr{velocity resolution} of 
3.75\,km\,s$^{-1}$.  The slit, \gr{of length $15^{\prime\prime}$}, was oriented at position angle 59\degr$\,$
 East of North (Figure~1) to cover the bright limbs of the
nebula in the radial direction.  A $K=11.3$~mag star (2MASSJ~21070267+4214099;
marked with a ``G'' in Figure~1), located 10\arcsec~east of the center
of the nebula, was used for guiding.  The offsets shown in Figure 1 are 
relative to the position of the central star determined\footnote{Although the $K \sim 14.3$ (Latter et al.\ 2000) central star was not apparent in the real-time image provided by the slit-viewing camera, it could be detected by co-averaging 600 camera frames, each of duration 1 sec,  obtained during the spectroscopic observations} on the $K$-band slit-viewing camera: 
\gr{$\alpha = \rm 21^h\,07^m\,1\fs 793, \delta=42\degr\,14^{\prime\prime}\,9\farcs 79$ (J2000).}

{The guide star was kept on the same pixel of the
detector on the slit-viewing camera throughout each
integration. Because the angular size of the nebula is comparable to
the slit length, blank sky emission could not be sampled by
nodding the telescope while keeping the science target inside
the slit.  We moved the slit to 16\arcsec~east of the \gr{center of the} nebula 
every 5 minutes to
record the sky background spectrum, while
keeping the guide star inside the field of view of the
camera. The seeing was superb on the night of July 13, 
consistently below 0\farcs4 at $K$ band, while on night of September 6th 
the seeing varied between $0\farcs3$ and $0\farcs6$.}

The total on-source
integration time with the {\tt Lp1} setting was 106 minutes (not
including the time for sampling the sky emission). The early-type
standard stars HR\,7001 (A0V) and HR\,7557 (A7V) were observed
every two hours with similar airmasses to that of NGC~7027. Flat fields
were obtained using the calibration unit of the instrument prior to 
every change in the telescope pointing.  Weather conditions
were not optimal on the night of September 6, with high
humidity and high cirrus clouds.  The total on-source time for {\tt Lp2}
was 66 minutes. The standard star HR\,7557 was
observed before the science observations.

\section{Data Reduction}

Data reduction was performed using the program suite {\it
  Spextool} ver. 5.0.2 (Cushing et al. 2004) adapted for the
iSHELL data.  {\it Spextool} coadds the raw frames, subtracts
the spectrogram image of the offset sky, normalizes the pixel
responses by dividing the spectrogram by the flat-field images,
calculates the wavelength mapping on the detector with reference to
atmospheric emission lines, and extracts one dimensional
spectra from the given apertures. In order to measure the spatial
variation of the emission lines, \gr{15} extraction regions were
defined along the slit length
(Figure~1). The size and the separation of these regions were each
1\arcsec.  Telluric lines
were removed by dividing the science spectra by the spectra of
the standard stars using the {\it xtellcor} code (Vacca et
al. 2003) that is part of {\it Spextool}. The broad \ion{H}{1}
recombination lines of the early-type standard stars were
removed at the same time. The one dimensional spectra consist
of 17 strips each representing one diffraction order. The spectral
strips were stitched together with {\it xmergeorders} within 
the {\it Spextool} suite.


The observed spectra of the standard star HR\,7557 were used to flux calibrate
the NGC 7027 spectra.  When narrow slit observations
of an extended source are flux calibrated
against a point source such as a standard star, it is necessary to account for
slit loss in the standard star observations.  The required slit loss 
correction was determined by a comparison of standard star spectra obtained 
with narrow (0\farcs375) and wide (4\farcs0) slits.


A $K$ band image of the nebula was constructed from the slit
viewing images. The astrometry was established referring to the
coordinates of three stars in the field, GAIA source ID
1969656406932827392 (the guide star in Figure~1),
1969656406922350848, and 1969656406922372480. The $K$ band image
was sliced along the putative slit position, and the cross-cut
of the image was cross-correlated with the spatial profile along
the slit of the two dimensional spectrogram in order to
calculate the accurate positions of the apertures on the image
(Figure~2).  This procedure indicated that the extraction region number X, 
counting from NE to SW as indicated in Figure 1, is centered at offset $\Delta \theta = {\rm X} - 8.1875$ arcsec from the central star, where positive values of $\Delta \theta$ refer to offsets in the SW direction.

\begin{figure}
\includegraphics[scale=0.55,angle=-90]{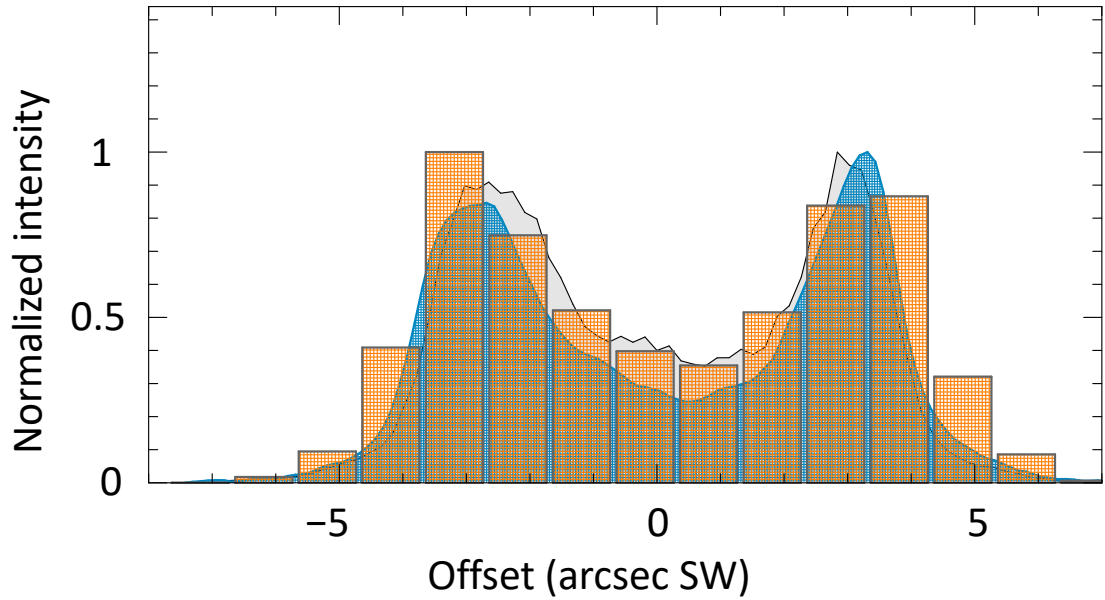}
\caption{Intensity profiles along the slit.  Blue: $L$-band continuum flux measured with the spectrograph (order
142 of Lp1, covering 3.6302 -- 3.6557 $\mu$m).  Orange: 3.52 $\mu$m continuum flux measured in each extraction region.  Grey: K-band intensity measured in the slit viewing camera.  Offsets 
are relative to the measured position of the central star, \gr{$\alpha = \rm 21^h\,07^m\,1\fs 793, \delta=42\degr\,14^{\prime\prime}\,9\farcs 79$ (J2000), with offsets to the SW defined as positive.}}
\end{figure} 

\section{Results}

With its large simultaneous spectral coverage and its high spectral resolution, the iSHELL spectrograph provides a large amount of data in a single setting.  In Appendix A, Figures A1 - A7, \gr{we present the complete spectrum obtained toward NGC 7027 and summed over the 11 central extraction regions (numbered 3 -- 13 in Figure 1)}.  As discussed in Appendix A, more than \gr{60} identifiable spectral lines are detected unequivocally: these include recombination lines of \ion{H}{1}, \ion{He}{1}, and \ion{He}{2}; vibrational lines of H$_2$ and HeH$^+$; a strong [\ion{Zn}{4}] fine structure line; and 9 rovibrational lines of CH$^+$.  The latter were detected for the first time in an astronomical source, following a recent measurement of their frequencies in the laboratory (Domenech et al.\ 2018), and will be analysed in a future publication. 
Figure 3 shows an expanded view of four spectral lines after subtraction of a linear baseline: the \ion{H}{1} $19-6$, \ion{He}{2} $13-9$, \ion{He}{1} $5^3D-4^3P^o$, and HeH$^+$ $v=1-0\, P(1)$ lines at 3.64593, 3.54431, 3.70357, and $3.51629\,\mu$m, respectively.   

The spectra shown in Figure 3 {were obtained from the sum of extraction regions \gr{5, 6, 11, and 12}, corresponding to offsets in the intervals $\Delta \theta = [-3.7, -1.7]$ and $\Delta \theta = [+2.3,+4.3]$}.  These extraction regions cover the bright limbs on either side of the central source (see Fig.\ 1). Solid lines show Gaussian fits to the line profiles, obtained using the scipy.optimize.curve$\_$fit routine in  python to fit the lines using the Levenberg-Marquardt algorithm. The line fluxes, widths, and centroids were allowed to vary freely and independently for each line, and are tabulated in Table 1 with their standard errors.   The fit to the HeH$^+$ $v=1-0\,P(1)$ line has a peak intensity of \gr{${(2.33 \pm 0.17)} \times 10^{-15} \rm \, W \, m^{-2} \mu m^{-1}$}, implying a detection at the $\gr{14}\,\sigma$ level.

\begin{figure}
\includegraphics[scale=0.65,angle=0]{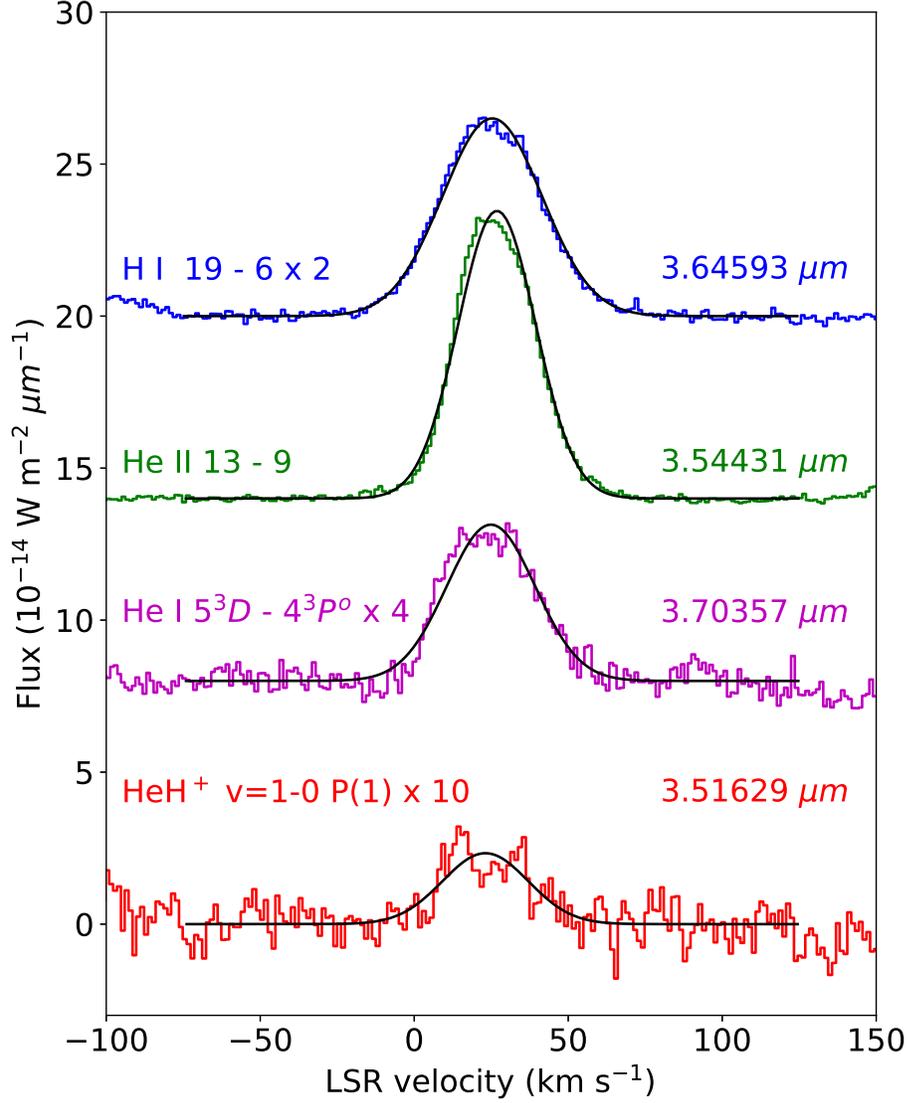}
\caption{Spectra of the \ion{H}{1} $19-6$, \ion{He}{2} $13-9$, \ion{He}{1} $5^3D-4^3P^o$, and HeH$^+$ $v=1-0\, P(1)$ lines at 
3.64593, 3.54431, 3.70357, and $3.51629\,\mu$m, respectively, after subtraction of a linear baseline (and the introduction
of vertical offsets for clarity).  These spectra were obtained from the sum of extraction regions \gr{5, 6, 11, and 12}, corresponding to offsets in the intervals $\Delta \theta = [-3.7, -1.7]$ and $\Delta \theta = [+2.3,+4.3]$.}
\end{figure}

\begin{deluxetable}{lccccc}
\tabletypesize{\footnotesize}
\tablecaption{Gaussian fit parameters for the lines plotted in Figures 3 and 4}

\tablehead{
Line & Rest wavelength & Peak flux$^a$ & Integrated flux & $v_{\rm LSR}^b$ & FWHM$^c$ \\
& ($\mu$m) 	   &  &  ($\rm 10^{-18}\,W\,m^{-2}$) & 
	   ($\rm km\,s^{-1}$) & ($\rm km\,s^{-1}$) \\}
\startdata
\ion{H}{1} $19-6$ 		& 3.64593 & $32.5 \pm 0.19^d$ & $16.1 \pm 0.12$  & $25.3 \pm 0.11$ & $38.0 \pm 0.28$\\
\ion{He}{2} $13-9$ 		& 3.54431 & $94.6 \pm 0.58$   & $35.5 \pm 0.26$  & $26.9 \pm 0.09$ & $29.6 \pm 0.22$\\
\ion{He}{1} $5^3D - 4^3P^0$	& 3.70357 & $12.9 \pm 0.28$   & $5.74 \pm 0.15$  & $24.9 \pm 0.35$ & $33.8 \pm 0.9$\\
$\rm HeH^+ \, v=1-0\,P(1)$ 		& 3.51629 & $2.33 \pm 0.17$   & $0.96 \pm 0.09$  & $23.1 \pm 1.15$ & $32.8 \pm 2.9$\\
$\rm HeH^+ \, v=1-0\,P(2)$ 		& 3.60776 & $2.56 \pm 0.34$   & $1.08 \pm 0.14$  & Note (e)  & Note (e) \\
\enddata
\tablenotetext{a}{Units of ($\rm 10^{-15}\,W\,m^{-2}\,\mu m^{-1}$)}
\tablenotetext{b}{Centroid velocity with respect to the local standard of rest}
\tablenotetext{c}{Full width at half maximum}
\tablenotetext{d}{Standard errors ($1\,\sigma$ statistical errors only)}
\tablenotetext{e}{Parameters fixed to those obtained for $\rm HeH^+ \, v=1-0\,P(1)$}

\end{deluxetable}

Based on the excitation model of HeH$^+$ described in Section 5.2 below, we expect the \gr{v$=1-0\, P(1)$ and 
$P(2)$} transitions to be
the strongest HeH$^+$ rovibrational lines in the bandpass.  
\gr{All other lines are predicted to be weaker by a factor of $\sim 3$ or more.}  This behavior results from
the facts that (1) at the densities of relevance here, most HeH$^+$ molecules are in the ground $(v,J)=(0,0)$ state;
and (2) electron-impact excitation from $(v,J)=(0,0)$ to $(v^\prime,J^\prime)=(1,0)$ and $(1,1)$ is strongly favored 
over excitation to $v^\prime=1$ states with $J^\prime > 1$, according to cross-sections computed by {\v{C}}ur{\'\i}k \& Greene (2017).

The $v=1-0\,P(2)$ line lies close to a strong recombination line ($n=20-6$) of atomic hydrogen.  
In Figure 4, we show the line profile obtained for 
the \ion{H}{1} $20-6$ line (blue histogram), together with that for \ion{H}{1} $19-6$ (black histogram).  
Here, the \ion{H}{1} $19-6$ line has been scaled by
a factor 0.875, as needed to match the peak of the \ion{H}{1} $20-6$ line.  As in Figure 3, 
these spectra were obtained for the sum of 
extraction regions \gr{5, 6, 11, and 12} as in Figure 3.  The \ion{H}{1} $20-6$ observations were affected by
three narrow atmospheric absorption features within the spectral range that is shown here: the affected spectral channels 
have been excised, as indicated by the gaps in the spectra.
Relative to the the \ion{H}{1} $19-6$ line, which is very nearly Gaussian (Figure 3),
a clear excess is observed in the red wing of the \ion{H}{1} $20-6$ line.  The red histogram shows this excess,
 with the velocity scale now shifted as appropriate for the HeH$^+$ $v=1-0\,P(2)$ line wavelength.  {The regions affected by three narrow atmospheric absorption were similarly excised from the HeH$^+$ $v=1-0\,P(2)$ spectrum (where they appear as gaps that are shifted relative to those in the blue histogram because of the different rest wavelength).}

\begin{figure}
\includegraphics[scale=0.65,angle=0]{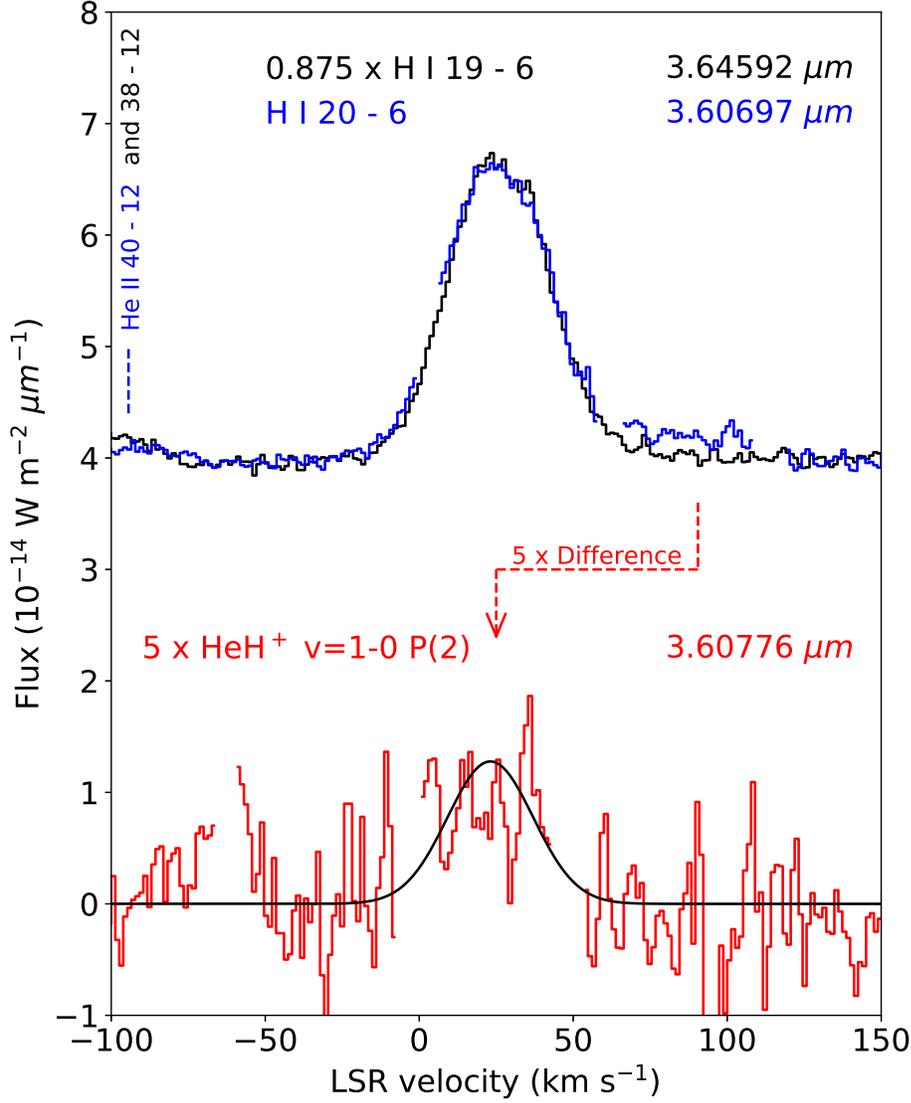}
\caption{Blue histogram: spectrum of the \ion{H}{1} $20-6$ line, after subtraction of a linear baseline (and the introduction of a vertical offset for clarity).  
Black histogram: spectrum of the \ion{H}{1} $19-6$ line.
Red histogram: excess in the red wing of the \ion{H}{1} $20-6$ line, with the velocity scale shifted as 
appropriate for the HeH$^+$ $v=1-0\,P(2)$ line wavelength.  The black curve shows a Gaussian fit to this excess, 
with the line width and centroid again equal to those measured for the $v=1-0\,P(1)$ line.  
The extraction region is the same as for Figure 3.}
\end{figure}
   
The solid line shows a Gaussian fit to these residuals, with the line width and centroid set
equal to \gr{the values} measured for the $v=1-0\,P(1)$ line; the HeH$^+$ $v=1-0\,P(2)$ line is \gr{detected} at the \gr{8} $\sigma$ level.  
{As expected, no other rovibrational transitions of HeH$^+$
were detected.}

\begin{figure}
\includegraphics[scale=0.75,angle=0]{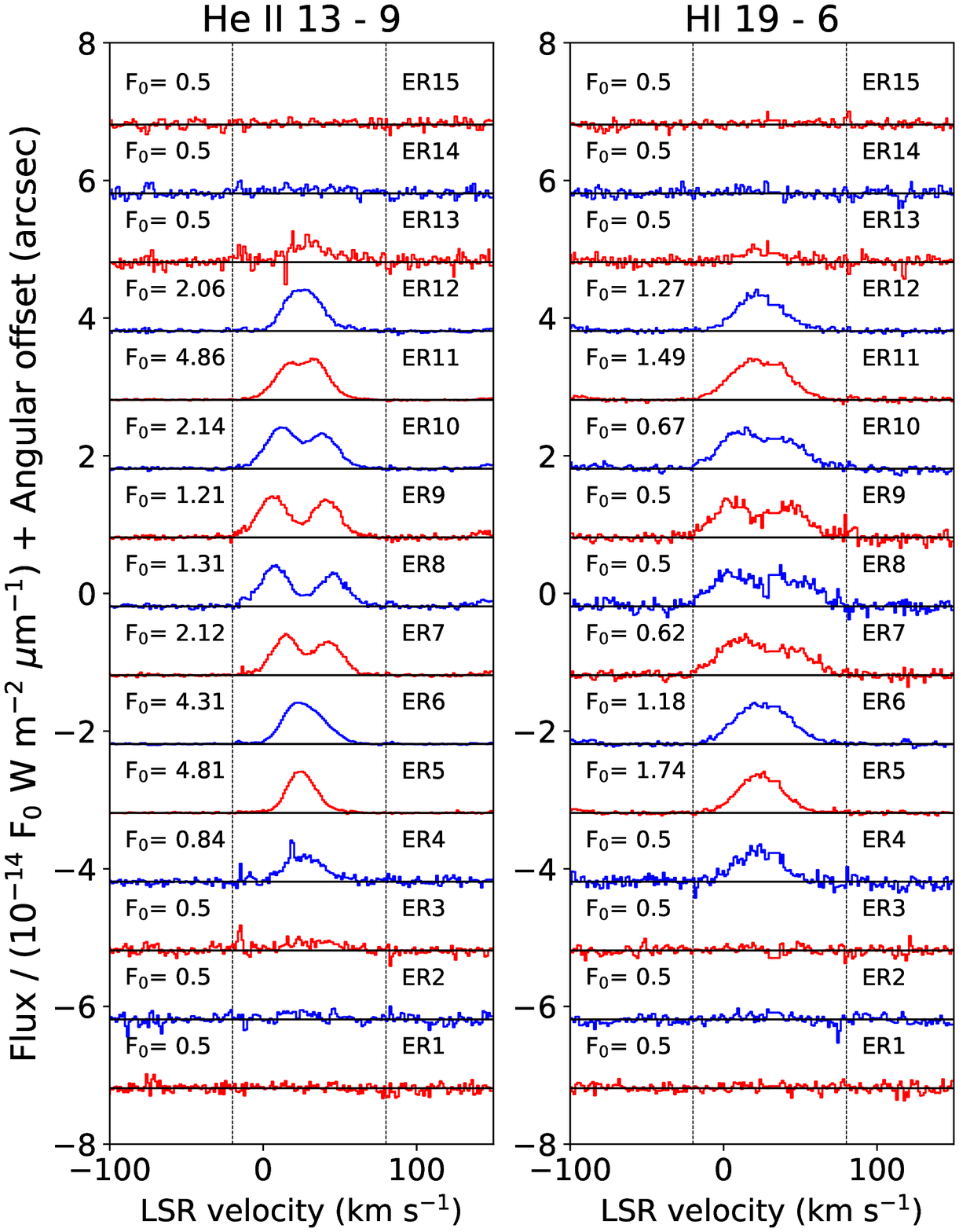}
\caption{Individual spectra obtained in each extraction region for the  \ion{He}{2} $13-9$ (left panel) and \ion{H}{1} $19-6$ (right panel) lines.  Note the different scaling factors (F$_0$) adopted for each region.  \gr{The alternating red and blue colors for the spectra are merely introduced for clarity}.  \gr{The numbering of the extraction regions (ER) 
is defined in Figure 1.}}
\end{figure}

\begin{figure}
\includegraphics[scale=0.75,angle=0]{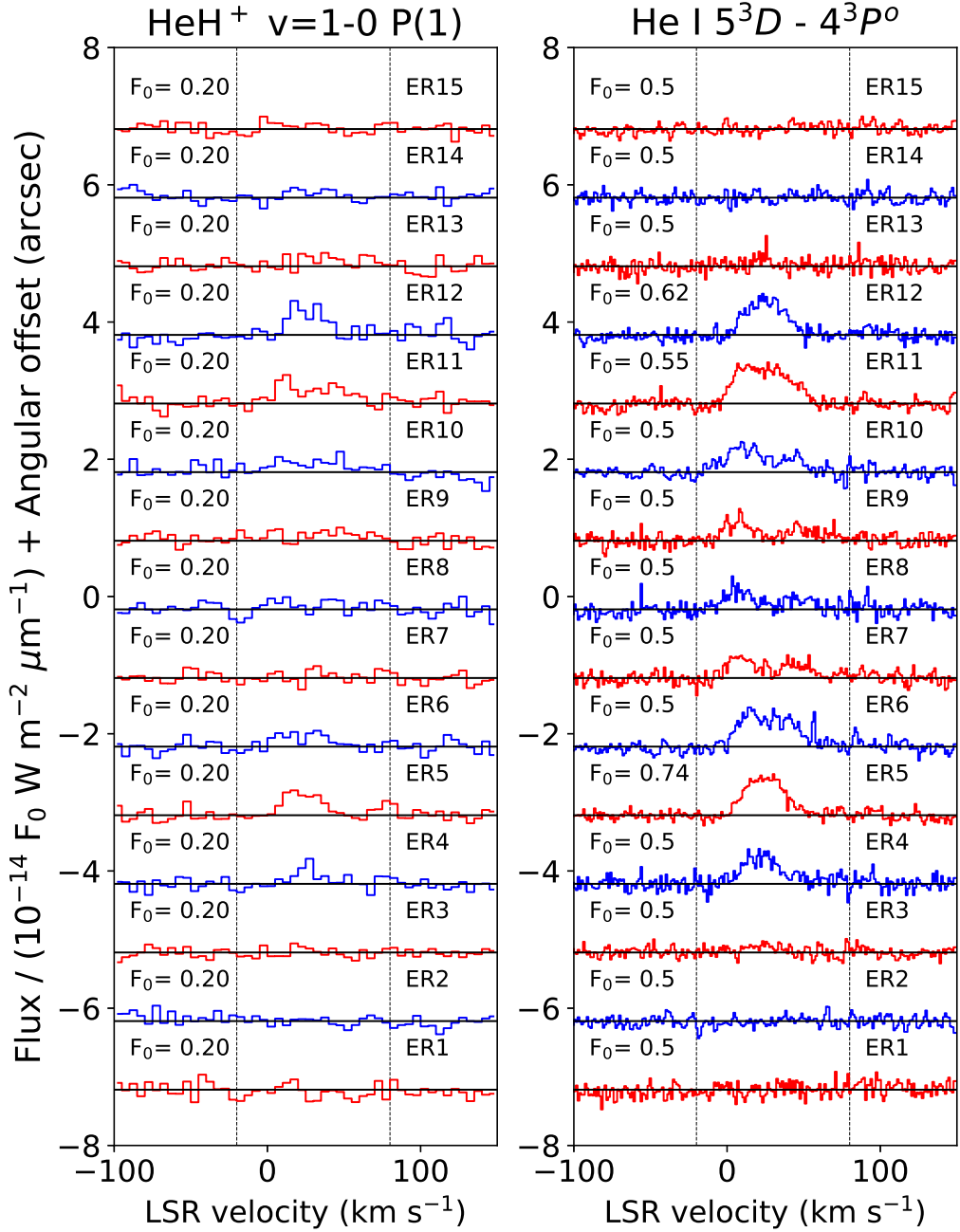}
\caption{Same as Figure 5, but for the  HeH$^+$ $v=1-0\, P(1)$ (left panel) and \ion{He}{1} $5^3D-4^3P^o$ (right panel) lines.  The HeH$^+$ $v=1-0\, P(1)$ 
spectra, which are individually of low SNR, have been rebinned to a channel width of $5\,\rm km\,s^{-1}$.}
\end{figure}

\begin{figure}
\includegraphics[scale=0.75,angle=0]{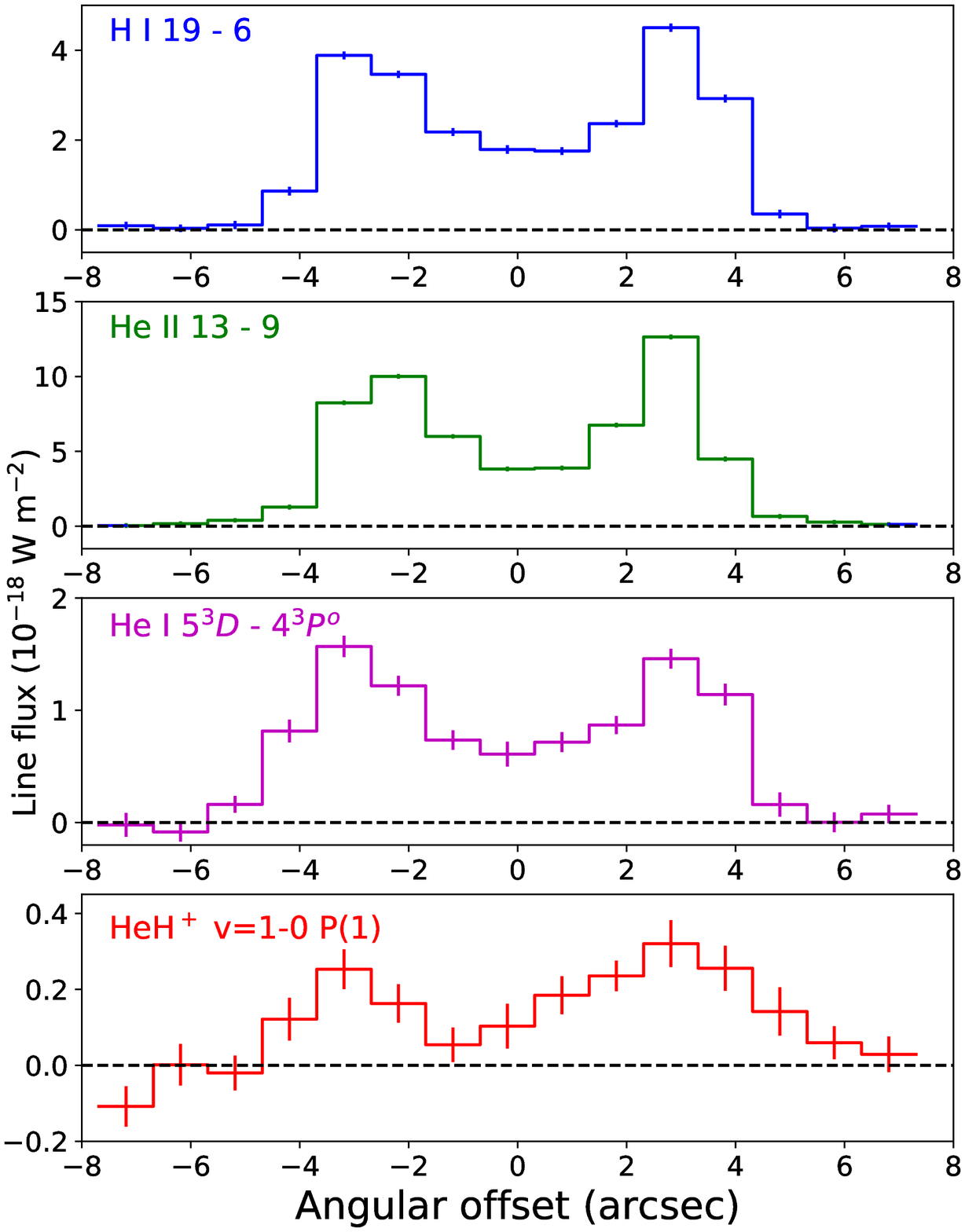}
\caption{Spatial variation of line fluxes along the slit, with 1$\sigma$ error bars.
The line fluxes are those measured in each $0\farcs375 \times 1\farcs0$ extraction region.}
\end{figure}

In Figures 5 and 6, we show the individual spectra obtained for each extraction region.  Within the part of the slit that covers the nebula, the \ion{H}{1}, \ion{He}{2} and \ion{He}{1} recombination lines are still detected at high signal-to-noise ratios (SNR) within the individual extraction regions of length $1\arcsec$.  As expected for an expanding shell, the line profiles become double-peaked near the center \gr{of} the nebula.  {The spectral lines were integrated over the --20 to 80\,$\rm km\,s^{-1}$ LSR velocity range to obtain the line flux for each extraction region.  Figure 7 shows the velocity-integrated line fluxes obtained, as a function of position along the slit.}

\section{Discussion}

\subsection{Model for HeH$^+$ in NGC 7027}

\begin{deluxetable}{ll}
\tablecaption{Parameters adopted in the source model$^a$}

\tablehead{Parameter & Value}
\startdata
Distance & 980 pc \\
Inner angular radius & \gr{$3\farcs 1$} $^b$ \\
Outer angular radius & \gr{$4\farcs 6$} $^b$ \\
Stellar luminosity & $1.0 \times 10^4\,L_{\odot}$  \\
Stellar effective temperature & $1.90 \times 10^5\,\rm K$  \\
Pressure/k & $1.6 \times 10^9\, \rm K\, cm^{-3}$ (constant)$^c$ \\
Helium abundance & 0.120 \\

\enddata
\tablenotetext{a}{Except where noted, the parameters adopted are those given by Zijlstra et al. (2008)}
\tablenotetext{b}{Geometric mean of the semi-major and semi-minor axes of the photoionized region}
\tablenotetext{c}{Adjusted to match the outer angular radius}

\end{deluxetable}

\begin{deluxetable}{llll}
\tabletypesize{\footnotesize}
\tablecaption{Reaction rates adopted in the chemical model$^a$}

\tablehead{Reaction && Rate coefficient & Reference/note \\
				&& ($\rm cm^3 \, s^{-1}$) \\}
\startdata
$\bf (R1)$ & $\bf He^+ + H \rightarrow HeH^+ + h \nu$  		&  $1.44 \times 10^{-16}$ & (b,c) \\
$\bf (R2)$ & $\bf He(2^3S) + H \rightarrow HeH^+ + e$  		&  $5.4 \times 10^{-11} T_4^{-0.5}$ & Waibel et al.\ (1988)$^c$\\
$\bf (R3)$ & $\bf HeH^+ +e \rightarrow He + H$         		&  $4.3 \times 10^{-10} T_4^{-0.5}$ & Novotn{\'y} et al.\ (2019) \\
$\bf (R4)$ & $\bf HeH^+ + H \rightarrow H_2^+ + He$    		&  $1.7 \times 10^{-9}$ & Esposito et al.\ (2015) \\
$(R5)$ & $\rm He + H^+ \rightarrow HeH^+ + h \nu$       &  $5.6 \times 10^{-21} T_4^{-1.25}$ & Fit to Jurek et al.\ (1995) \\
$(R6)$ & $\rm He(2^3S) + H^+ \rightarrow HeH^+ + h \nu$   &  $3.6 \times 10^{-17} T_4^{-0.7}$ & Fit to Loreau et al. (2013) \\
$(R7)$ & $\rm He^+ + H^- \rightarrow HeH^+ + e$         &  $3.2 \times 10^{-11} T_4^{-0.34}$ & Le Padellec et al. (2017)$^c$ \\
$(R8a)$ & $\rm H^+ + H \rightarrow H_2^+ + h \nu$   	&  $2.3 \times 10^{-16} T_4^{1.5}$ & Fit to Ramaker and Peak (1976)\\
$(R8b)$ & $\rm H_2^+ + He \rightarrow HeH^+ + H$     	&  $3 \times 10^{-10} \exp(-0.6717/T_4)$ & Black (1978)\\ 
$(R8c)$ & $\rm H_2^+ + H \rightarrow H_2 + H^+$   		&  $6.4 \times 10^{-10}$ & Karpas et al.\ (1979)\\
$(R8d)$ & $\rm H_2^+ + e \rightarrow H + H$   		 	&  $3 \times 10^{-9}\ T_4^{-0.4}$ & Schneider et al. 1994, 1997$^d$\\
$(R9)$ & $\rm HeH^+ + h \nu \rightarrow He^+ + H$     	&  (see text) & Cross-section from Miyake et al. (2011)  \\

\enddata
\tablenotetext{a}{Only the four reactions in boldface type are found to make a significant contribution}
\tablenotetext{b}{Based on the cross-section presented by Vranckx et al. (2013).
The published cross-sections are for collisions in which He$^+$ (1s) and H (1s) 
are in the singlet state with total spin 0 (Loreau 2019, private communication), 
but as only one-quarter of collisions will have spin 0, 
the rate has been reduced by a factor of four.}
\tablenotetext{c}{The primary reference gives the cross-section as a function of
collision energy, which we used to obtain a rate coefficient for a Maxwell-Boltzmann distribution of particle velocities}
\tablenotetext{d}{Fit obtained by Stancil et al. (1998)}
\end{deluxetable}

To interpret the observational results described above, we have used a model for NGC 7027 similar to that discussed in G19.  The model parameters are given in Table 2.  We used the CLOUDY photoionization code
to model the temperature and densities of H, H$^+$, H$^-$, He, He($2^3S$), He$^+$, He$^{++}$ and electrons as a function of position in the nebula \gr{down to a temperature of 800~K.  
We then} computed the equilibrium abundance of HeH$^+$ given the formation and destruction processes listed in Table 3. 
{As in G19, our standard model was computed under the assumption of constant pressure.  With an alternative assumption of constant density, we found that the H nucleus density needed to match the observed Str\"omgren radius was $n_{\rm H} = 4.37 \times 10^4\,\rm cm^{-3}$.  The predicted line fluxes were not greatly altered: the \ion{H}{1}, \ion{He}{2}, \ion{He}{1}, and HeH$^+$ line flux predictions changed by $+6 \%$, $+7 \%$, $+3 \%$ and $-26 \%$.  The decrease in the predicted HeH$^+$ line flux reflects the fact the HeH$^+$ emission peaks within a shell near the Str\"omgren radius; because this region is somewhat cooler than the fully ionized region, the density is lower in the constant density model than in the constant pressure model.  Because the sound crossing time is comparable to the age of the system, it is not clear which model is more appropriate.}

Several changes to the chemistry from the analysis of G19 have been implemented as described below.  For the rate coefficient for dissociative recombination (DR) of HeH$^+$, which is a dominant destruction process, we now adopt the value obtained very recently from an experiment at the Cryogenic Storage Ring (CSR) in Heidelberg (Novotn{\'y} et al.\ 2019).  Although at low temperatures the DR rate was found to vary strongly with the rotational state of HeH$^+$, the value at the temperatures ($\sim 10^4\,$K) of present interest is independent of $J$.  It is, however, somewhat higher ($4.3 \pm 0.9 \times 10^{-10}\,\rm cm^{3}\,s^{-1}$) than that adopted by G19.  That value, $3 \times 10^{-10}\,\rm cm^{3}\, s^{-1}$, was based on an earlier experiment by Str\"omholm et al.\ (1996) after correction of an error in the original paper.  (The two experimental results are consistent with each other given the likely systematic errors.)  For the destruction of HeH$^+$ in reaction with H, we adopt results obtained in the more recent calculations of Esposito et al.\ (2015), which are somewhat larger ({by a factor $\sim 1.4$}) than those (Bovino et al.\ 2011) we adopted previously.  {In addition, we have included the formation of HeH$^+$ by the associative ionization of atomic H with He in the metastable 2$^3S$ state [reaction $(R4)$].  Based on more recent measurements (Waibel et al.\ 1988) of the cross-section for reaction $(R2)$, we obtained a rate coefficient \gr{five times} as large as that estimated by Roberge \& Dalgarno (1982); with this modification, the associative ionization reaction $(R2)$ becomes a significant source of HeH$^+$.}  

{For completeness, we have also included four additional HeH$^+$ formation pathways and one additional destruction process.  These are (1) formation of HeH$^+$ by radiative association of He with H$^+$ [reaction $(R5)$]; (2) formation of HeH$^+$ by radiative association of He(2$^3$S) with H$^+$ [reaction $(R6)$)]; formation of HeH$^+$ by an associative ionization reaction of He$^+$ with H$^-$ [reaction $(R7)$)]; (4) formation of H$_2^+$ via radiative association of H and H$^+$ [reaction $(R8a)$], followed by proton transfer to He [reaction $(R8b)$]; the efficiency of the latter pathway is reduced by reactions $(R8c)$ and $(R8d)$ that destroy H$_2^+$; and (5) photodissociation of HeH$^+$ [reaction $(R9)$], the rate of which is determined by the ultraviolet radiation field and was computed at each point in the nebula.}

\begin{figure}
\includegraphics[scale=0.72,angle=0]{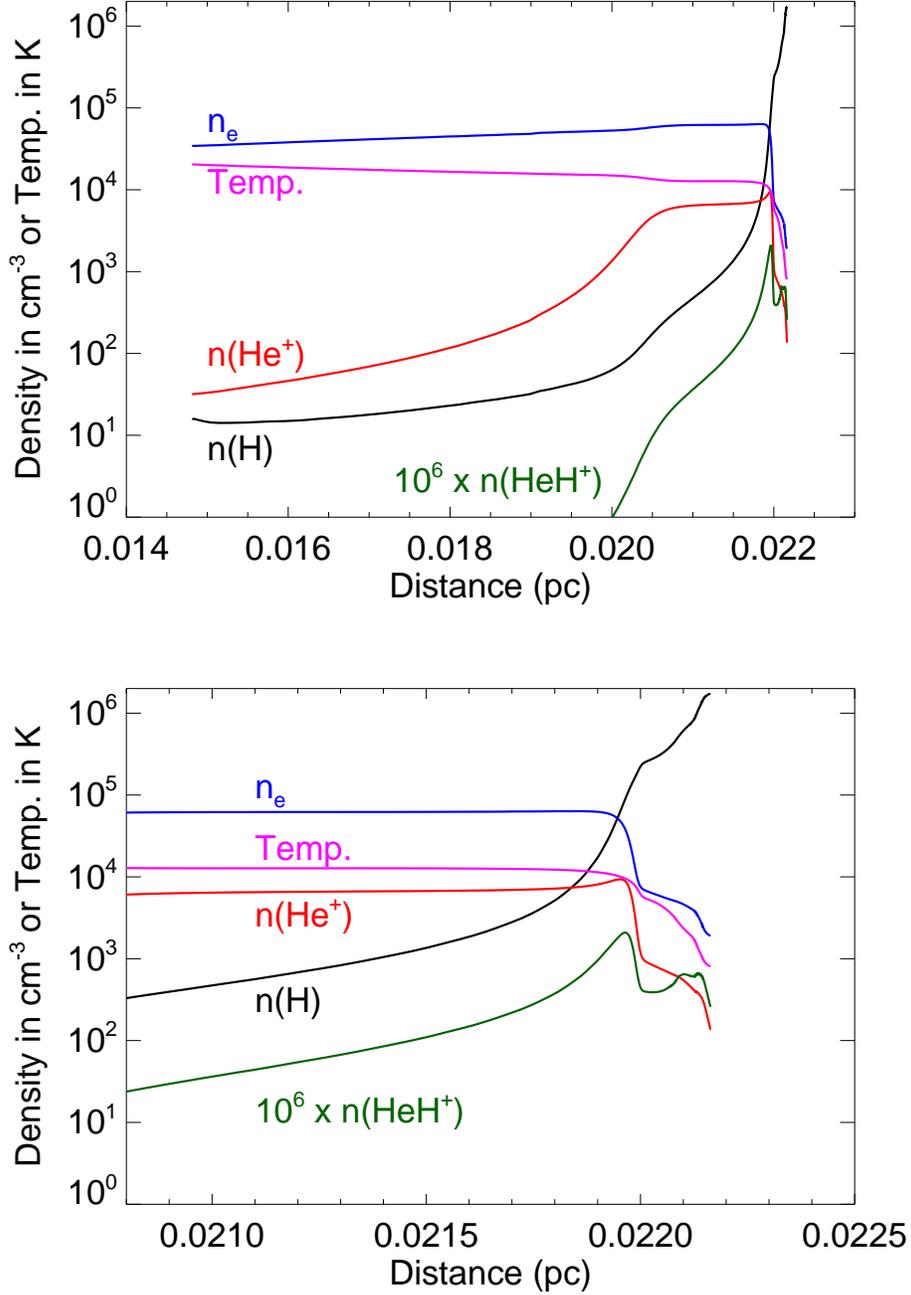}
\caption{Top panel: temperature profile and the density profiles for several key species.
Bottom panel: zoomed view of top panel in the region close to the Str\"omgren radius.}
\end{figure}

\begin{figure}
\includegraphics[scale=0.72,angle=0]{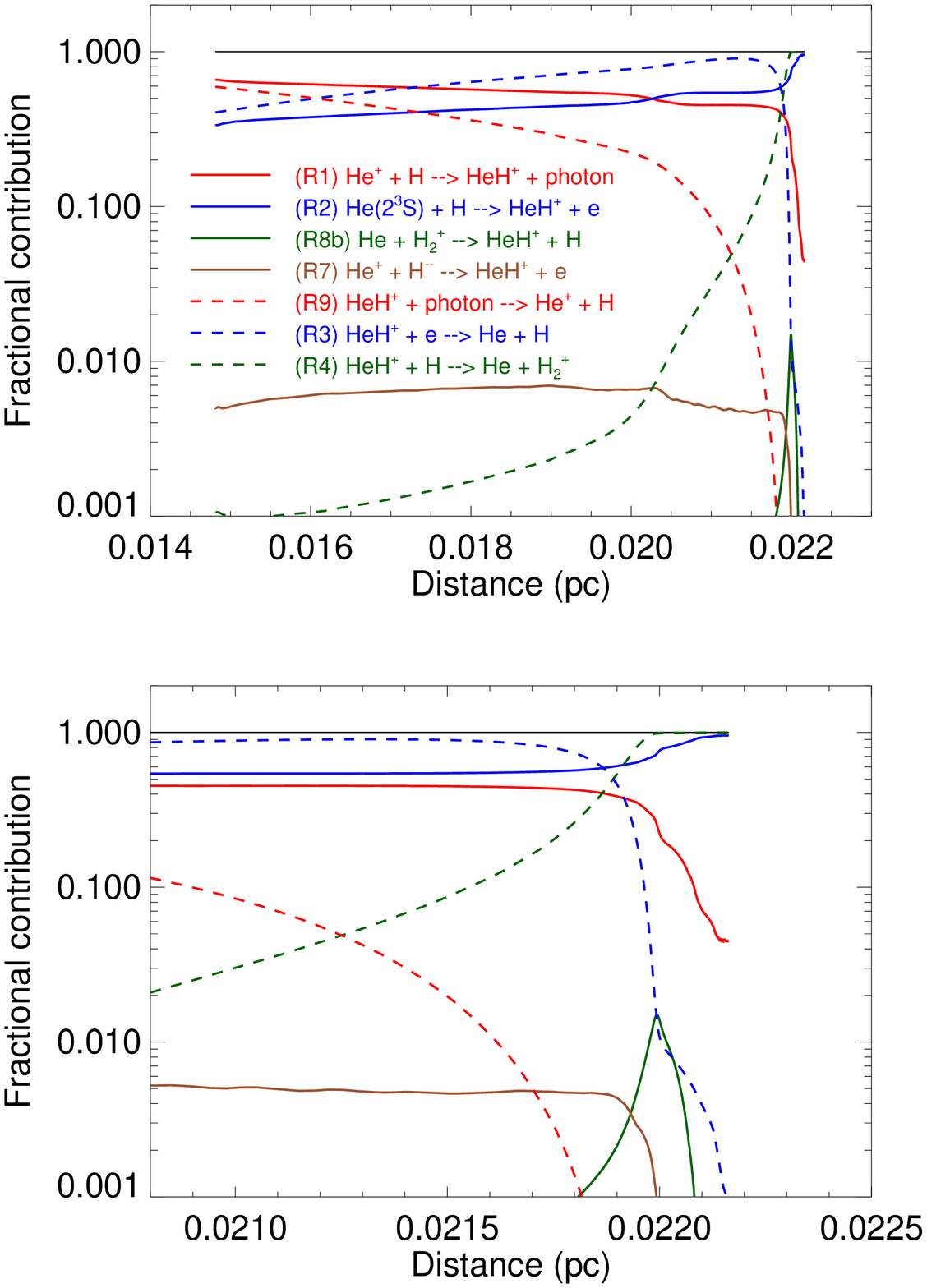}
\caption{Top panel: fractional contribution of various processes to the formation and destruction of HeH$^+$, with formation process plotted with solid lines and destruction processes with dashed lines.
The chemistry is dominated by just four reactions: $(R1)$, $(R2)$, $(R3)$, and $(R4)$.
Bottom panel: zoomed view of top panel in the region close to the Str\"omgren radius.}
\end{figure}

The combined effect of these changes is to increase the predicted HeH$^+$ emission flux by a factor $\sim {1.3}$.
Figure 8 shows the temperature and the density profiles for several key species (top panel with a zoomed view in the middle panel), now computed with these modifications to the HeH$^+$ chemistry, while Figure 9 shows the fractional contributions of the various destruction and formation processes for HeH$^+$.  Reactions $(R5)$ and $(R6)$ never make a fractional contribution to the HeH$^+$ formation rate that exceeds $10^{-3}$; they therefore fail to appear in Figure 9.

\subsection{Comparison of the observed line intensities with the model}

The predicted fluxes for the strongest observed recombination lines of He$^+$, He, and H are given in Table 4, along with their values measured in a $0\farcs375 \times 11^{\prime\prime}$ region spanning the minor axis of the nebula 
(extraction regions 3 -- 13 in Figure 1).  {Here, in contrast to Table 3,
the measured values were obtained from direct integration over the --20 to 80\,$\rm km\,s^{-1}$ LSR velocity range, without any assumption that the line profile is Gaussian.  The LSR velocity ranges --75 to --20 and 80 to 125\, $\rm km\,s^{-1}$ were used to define the continuum.} 
The predictions for He$^+$ and H were based on the Case B recombination line emissivities presented by Storey \& Hummer (1995).  In the case of the He recombination line, we made use of the emissivities computed by Bauman et al.\ (2005).    For the recombination lines, the {observed values are a factor $0.50 - 0.69$  times the model predictions.  Given the uncertainties in the flux calibration and the approximations made in the model (most notably the assumption of spherical symmetry), this may be regarded as acceptable agreement.}  

\begin{deluxetable}{lccccc}
\tabletypesize{\footnotesize}
\tablecaption{Comparison between the observed and predicted line fluxes}

\tablehead{
Line & Rest wavelength & Observed line flux$^a$ & Predicted line flux & Ratio \\
& ($\mu$m) 	   & ($\rm 10^{-18}\,W\,m^{-2}$) &  ($\rm 10^{-18}\,W\,m^{-2}$) & (observed/predicted) \\}
\startdata
\ion{H}{1} $19-6$ 			   & 3.64593 & \phantom{$^b$} $ 23.9\pm  0.22$ $^b$ 	& 34.6  & $0.69 \pm 0.006$ \\
\ion{H}{2} $13-9$ 			   & 3.54431 & $ 53.7\pm  0.21$	   				& 108   & $0.50 \pm 0.002$ \\
\ion{He}{1} $5^3D - 4^3P^0$	       & 3.70357 & $8.44 \pm 0.48$    					& 12.8  & $0.66 \pm 0.038$ \\
$\rm HeH^+ \, v=1-0\,P(1)$ & 3.51629 & $1.55 \pm 0.16$     					& 1.68  & $0.92 \pm 0.095$  \\
$\rm HeH^+ \, v=1-0\,P(2)$ & 3.60776 & $2.08 \pm 0.31$    					& 0.89  & $2.33 \pm 0.35$ \\
HeH$^+$ $J=1-0$	  	  & 149.137 & \phantom{$^b$}$163 \pm 32$ $^c$		& 56 $^c$ & $2.86 \pm 0.56$\\
\enddata
\tablenotetext{a}{Except for the HeH$^+$ $J=1-0$ line (see note c), the 
flux was measured in a $0\farcs375 \times 11^{\prime\prime}$ region spanning the minor axis of the nebula 
(extraction regions 3 -- 13 in Figure 1), }
\tablenotetext{b}{$1 \sigma$ statistical errors}
\tablenotetext{c}{Equivalent point source flux within a Gaussian beam of half-power-beam-width 14\farcs3.
The corresponding integrated main beam brightness temperature for the GREAT observations performed by G19
was $3.6 \pm 0.7 \,\rm K\, km \, s^{-1}$.  (The actual flux from the nebula is somewhat larger than the value tabulated
because the source is slightly extended relative to the SOFIA beam).}

\end{deluxetable}

Table 4 also lists predictions for the HeH$^+$ vibrational line fluxes and the integrated brightness temperature of the HeH$^+$ $J=1-0$ pure rotational transition observed with SOFIA/GREAT (G19), together with the measured values. 
\gr{Given the high fractional ionization within the emission region, collisional excitation by electrons is expected to dominate excitation by any other collision partner.  Moreover, formation pumping is expected to be relatively unimportant for states that are efficiently {excited} by collisions with electrons.}\footnote{\gr{This follows from the facts
that (1) within the HeH$^+$ emission region, the formation rate of HeH$^+$ is comparable to its destruction rate via dissociative recombination; and (2) the rate coefficient for dissociative recombination of HeH$^+$ is typically smaller than those 
for electron-impact excitation (for the states that we observed and at the temperature in the emission region).}}
For the excitation of rotational states within the ground vibrational state, we
adopted the thermal rate coefficients for electron-impact excitation presented recently by Ayoub and Kokoouline 
(2019; hereafter AK19), which are in excellent agreement with those obtained in an
independent recent calculation by {\v{C}}ur{\'\i}k \& Greene (2017; hereafter CG17) over the more limited 
temperature range (up to 3000 K) for which CG17 presented results.  {Another recent calculation by Hamilton et al.\ (2016; hereafter H16) reported 
similar rate coefficients for most collisional-induced transitions, but a rate coefficient for the excitation from
$J=0$ to $J=1$ that was $\sim 3$ times as large as that computed by AK19 and CG17.  The calculations of CG19 and AK19
may be considered more reliable, however, because they accounted for the presence of rotational (and vibrational) resonances 
in the e-HeH$^+$ spectrum.} 

We solved the equations of statistical \gr{equilibrium}
for the lowest five rotational states to determine the relative level populations and the
emissivity in the $J=1-0$ pure 
rotational transition.  At the electron densities of relevance in NGC 7027 (see Figure 8), the rotational excitation
is predicted to be subthermal, with most HeH$^+$ molecules in the ground rotational state.  Because the rate coefficients
for rovibrational excitation are roughly two orders of magnitude smaller than those for pure rotational excitation, 
the rotational level populations are not significantly \gr{affected} by the effects of rovibrational excitation.
For rovibrational excitation,  we adopted the cross-sections calculated by CG17 (their Figure 7).  Here, the 
cross-sections presented in CG17 were supplemented by results for all collisionally-induced transitions 
$(v,J) =(0,J_l) \rightarrow (1,J_u)$ with $J_u \le 4$ and $J_l \le 4$ \gr{({\v{C}}ur{\'\i}k 2019, private communication)}.   We used these cross-sections to obtain thermal rate
coefficients for $(v,J) =(0,J_l) \rightarrow (1,J_u)$, under the assumption that the excitation cross-sections were 
inversely proportional to energy for energies greater than ($> 0.2$~eV) those computed by CG17.  Combined with our solution for
the rotational population in $v=0$, these rovibrational rate coefficients could then be used to determine the emissivities
for the vibrational transitions that we have observed.

As noted previously by G19, our model significantly underpredicts the brightness temperature of the HeH$^+$ $J=1-0$ pure
rotational line; with the modifications to the chemical network described above, the model predictions now lie a {factor $\sim 3$} below the observed value (somewhat smaller than the factor 4 discrepancy given by G19).  
While the measured
strength of the HeH$^+$ v$=1-0$ $P(2)$ rovibrational line 
exceeds the model prediction by a similar factor (2.3), the flux measured for the HeH$^+$ v$=1-0$ $P(1)$ line is in good agreement with 
our model prediction.  The observed
v$=1-0\,P(1)/P(2)$ line ratio of $0.75 \pm 0.14$ ($1\, \sigma$ statistical error) is significantly 
smaller than the value of 1.9 predicted by the model.
{Because the v$=1-0\,P(2)$ line lies in the wing of a much stronger \ion{H}{1} recombination line and is affected by narrow atmospheric absorption features, there may
be significant systematic uncertainties in our determination of the line ratio.  Nevertheless, a discrepancy of this magnitude likely points to inaccuracies in the rate coefficients for vibrational excitation that are adopted in the model.
The v$=1-0\,P(1)/P(2)$ ratio, 
in particular, is primarily determined by the relative collisional rate coefficients and depends only very weakly on 
the gas temperature and density.}

For all collisionally-induced 
transitions within the ground vibrational state, AK19, CG17, H16 and Rabadan et al.\ 1998 obtained excitation
rate coefficients for electron impact that were in good agreement with each other  {except for the transition from $J=0$ to $J=1$. For this transition, as discussed previously, the H16 results are larger by a factor 3; if we use those results in place of those given by AK17, 
the predicted $J=1-0$ line intensity increases by a factor $\sim 2$ and is in closer agreement with the observations.  However, for the reasons given above, the calculations presented by AK19 and CG17 may be considered more reliable from the molecular physics standpoint.} 
For collisionally-induced transitions from $\rm v=0$ to $\rm v=1$, the only rotationally-resolved calculations of
the excitation rates are those of CG17.  Further calculations of electron-impact excitation
cross-sections for HeH$^+$ could be very valuable in understanding the discrepancies discussed above, 
particularly for vibrational transitions at collision energies above 0.2~eV.

If we assume the rate coefficients for collisionally-induced transitions within the ground vibrational state
to be the most reliable, based on the excellent agreement between the independent calculations of CG17 and AK19, 
the flux measured for the $J=1-0$ pure rotational line {may suggest} that the chemical model underestimates the HeH$^+$ 
abundance by a factor $\sim 3$.  The two main destruction mechanisms [reactions $(R3)$ and $(R4)$] have each been the subject of two independent investigations that yielded similar results [experimental in the case of $(R3)$ and theoretical in the case of $(R4)$]. 
These considerations suggest that the rate coefficients for the formation reactions $(R1)$ and/or $(R2)$ have been underestimated,  or that some important HeH$^+$ formation or excitation mechanism has been overlooked.  {Alternatively, or in addition, it may reflect approximations in our (spherically-symmetric) physical model for the source, or uncertainties in the rate coefficients we adopted for rotational excitation of HeH$^+$ in collisions with electrons.}

\begin{figure}
\includegraphics[scale=0.75,angle=0]{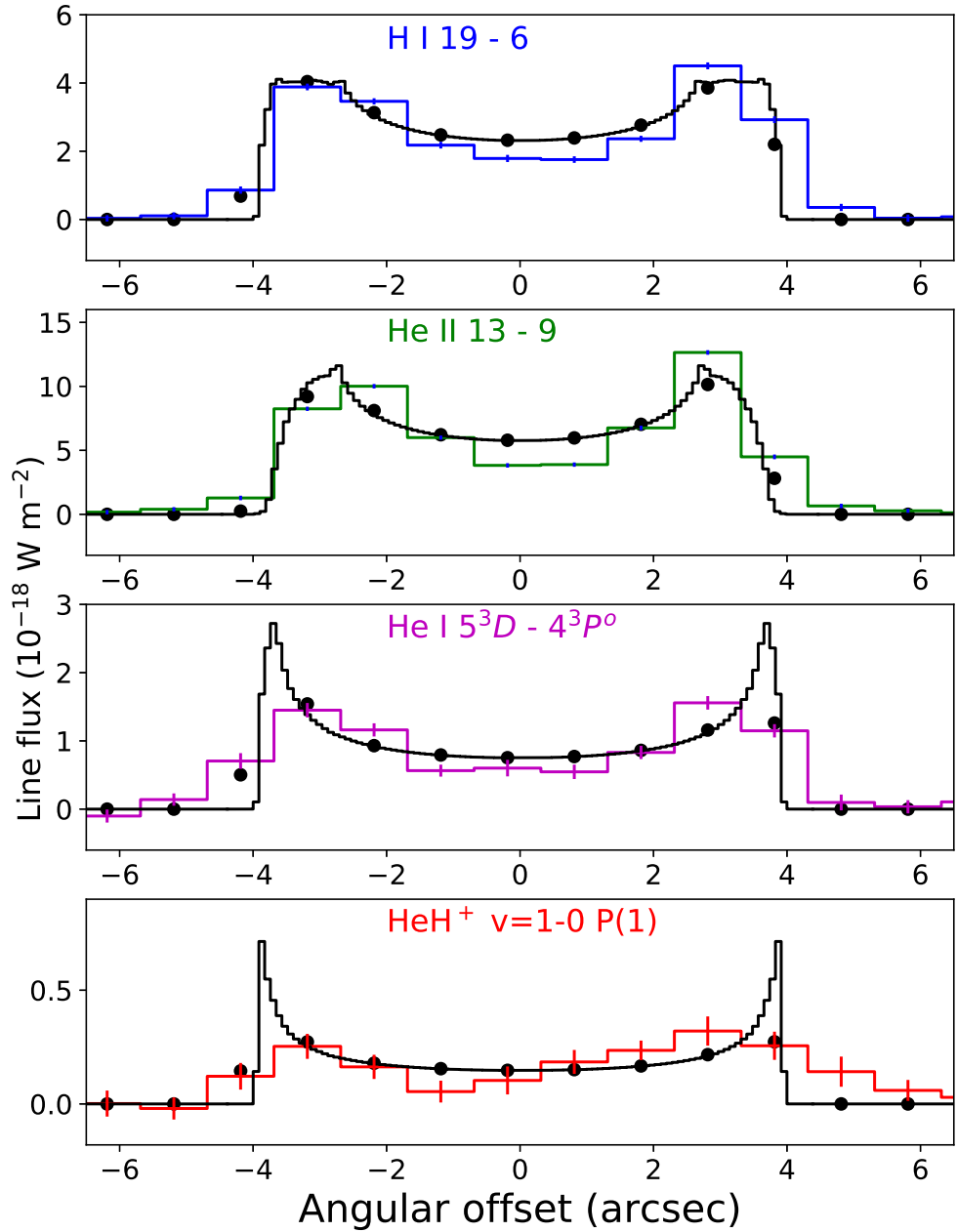}
\caption{Comparison of the observed spatial variation of line fluxes along the slit (colored histograms) with model predictions (black histogram).  The black points show the model predictions after rebinning onto the aperture extraction regions.}
\end{figure}

Based on our model for NGC 7027, we have also obtained predictions for the spatial variation of the emission line intensities along the slit.   {In Figure 10, we show how the intensities of several observed emission lines are predicted to vary with angular position.  As in Figure 7, the colored histograms show the observed spectral line fluxes as a function of position along the minor axis.  Because our model for the nebula makes the approximation of spherical symmetry and is adjusted to match the average of the semi-minor and semi-major axes, a scaling is needed to match the spatial profile along the minor axis.  Based on a fit to the strongest line (\ion{He}{2} $13 - 9$) we observed, we determined that scaling the predicted offsets by a factor of 0.86 was optimal.  Because the purpose of Figure 10 is to investigate the spatial variation of the line intensities rather than their absolute values, we also scaled the predicted fluxes separately for each line to optimize the fit to the observed profiles.  The black histogram shows the predicted spatial profiles, after the horizontal compression and vertical scalings decribed above. 
The results plotted in the black histogram were averaged over \gr{$0 \farcs 1$} bins.  The HeH$^+$ line and the He recombination line are predicted to be the most strongly limb brightened, because HeH$^+$ and He$^+$ are only abundant near the outer edge of the nebula. To facilitate a comparison between the model predictions and the observed spatial profiles, we have rebinned the predictions onto the 
extraction regions to obtain the black points in Figure 10.} 

{In the case of the \ion{He}{2} and \ion{H}{1} recombination lines, both of which were observed with a high SNR, the model clearly overestimates the intensities at the center of the nebula (or, equivalently, underestimates the degree of limb brightening).  This behavior has been noted previously, e.g. in a study of radio continuum observations published by Masson (1986). In that study, models for a spherically-symmetric emitting shell were shown to be
incapable of fitting the width of the limb emission at the same time as reproducing the large degree of limb brightening; this inconsistency could be resolved by models in which the ionized shell was assumed to be elongated along the line-of-sight to the observer, such that the ultraviolet flux near the projected center of the nebula was diminished and the associated free-free radio emission reduced.  This effect may also explain why the total observed line fluxes for the recombination lines (Table 4) are somewhat smaller than the model predictions.  For the weaker \ion{He}{1} and HeH$^+$ 
emission lines, the spatial profiles are not measured with a sufficient SNR to allow strong conclusions about limb brightening to be reached.}

\section{Summary}

\noindent 1) We have \gr{detected emission in the 3.51629$\,\mu$m v$=1-0$ $P(1)$ and 3.60776$\,\mu$m v$=1-0$ $P(2)$ rovibrational lines of the helium hydride cation (HeH$^+$) toward the planetary nebula NGC 7027.} The v$=1-0$ $P(2)$ line lies in the wing of a much stronger H recombination line, \gr{but could be readily separated thanks to the 
high spectral resolution of iSHELL}.

\noindent 2) The flux measured for the HeH$^+$ v$=1-0$ $P(1)$ line is in good agreement with 
our model for the formation, destruction and excitation of HeH$^+$ in NGC 7027, while the measured
strength of the HeH$^+$ v$=1-0$ $P(2)$ rovibrational line
exceeds the model predictions by a factor of 2.3.  
{The HeH$^+$ v$=1-0$ $P(1)/P(2)$ line ratio predicted by our model is a factor $2.5 \pm 0.5$ ($1 \sigma$ statistical error) larger than the measured value.  
Because the v$=1-0\,P(2)$ line lies in the
wing of a much stronger \ion{H}{1} recombination line and is affected by narrow atmospheric absorption features, there may be significant systematic uncertainties in our determination of the line ratio.  Nevertheless, a discrepancy of this magnitude likely points to inaccuracies in the rate coefficients for vibrational excitation that are adopted in the model.}

{\noindent 3) Our model also underpredicts the $J=1-0$ pure rotational line strength (G\"usten et al.\ 2019) by a factor 2.9.}  This disagreement may  
suggest that the rate coefficients for one or both of the dominant HeH$^+$-forming reactions $(R1)$ and/or $(R2)$ have been underestimated, or that some important HeH$^+$ formation or excitation mechanism has been overlooked.  {Alternatively, or in addition, it may reflect approximations in our (spherically-symmetric) physical model for the source, or uncertainties in the rate coefficients we adopted for rotational excitation of HeH$^+$ in collisions with electrons.}

\noindent 4) Our observations of NGC 7027, performed with iSHELL and covering the 3.26 - 3.93$\,\mu$m spectral region, led to the detection of more than sixty spectral lines; in addition to the \gr{HeH$^+$} v$=1-0$ $P(1)$ and $P(2)$ lines, these \gr{include} multiple recombination lines of \ion{H}{1}, \ion{He}{1} and \ion{He}{2}, rovibrational emissions from H$_2$, a fine structure line of [\ion{Zn}{4}], and nine rovibrational emissions from CH$^+$.  The latter were detected for the first time in an astronomical source (although CH$^+$ pure rotational emissions
had been detected previously toward this source by Cernicharo et al.\ 1997).  The identification
of CH$+$ vibrational lines was made possible by recent laboratory spectroscopy 
performed by Domenech et al.\ (2018), and
the implications of their discovery will be considered in a future publication.
There are several emission lines that have not yet been identified, the two strongest of which are at wavelengths of 3.2817 and 3.8102$\, \mu$m. 

\noindent 5) The observations provide information about the spatial distribution of the various line emissions along the minor axis of NGC 7027.  As expected, all the spectral lines we have considered are limb brightened.  For recombination lines of \ion{H}{1}, and \ion{He}{2}, the degree of limb brightening is somewhat larger than the predictions obtained for a spherically-symmetric model of the nebula; as discussed in previous studies, this probably arises because the nebula is elongated along the line-of-sight.  The SNR obtained for the \ion{He}{1} and HeH$^+$ lines is not sufficient to constrain strongly the degree of limb brightening for those transitions.

\begin{acknowledgements}

\gr{The observations reported here were carried out at the Infrared Telescope Facility (IRTF), which is operated by the University of Hawaii under contract NNH14CK55B with the National Aeronautics and Space Administration.}
We are very grateful to the IRTF director, John Rayner, for making unallocated engineering time available for this project.  We thank the IRTF support astronomer, Adwin Boogert, and the telescope operators at the summit -- Brian Cabreira, Greg Osterman, and Dave Griep -- for the excellent support they provided to our observing program.  \gr{We are very grateful to Roman {\v{C}}ur{\'\i}k for providing 
the results of unpublished collisional excitation calculations, and to} Evelyne Roueff for pointing out that HeH$^+$ can be formed via an associative ionization reaction of H$^-$ and He$^+$.  We thank Roman {\v{C}}ur{\'\i}k, Chris Greene, and Slava Kokoouline
for useful discussions; and Adwin Boogert, Jean-Pierre Maillard, and John Rayner for helpful comments on the manuscript.  \gr{TRG's research is supported by the Gemini Observatory.} 
M.G. is supported by the German Research Foundation (DFG)
grant GO 1927/6-1.

\end{acknowledgements}

\vfill\eject
\centerline{\bf{Appendix A: Full 3.26 - 3.93$\,\mu$m spectrum obtained toward NGC 7027}} 

In Figures A1 -- A7, we present all the spectra obtained toward NGC 7027, with
labels indicating the \gr{62} spectral features that have been detected. \gr{These spectra were obtained by
summing over the 11 central extraction regions (numbered 3 -- 13 in Figure 1).  Spectral regions that were
severely affected by atmospheric absorption have been excised, resulting in the gaps that are
apparent (particularly in Figures A1 and A2)}.  \gr{In the case of the Lp2 spectra (Figures A5 -- A7), obtained under 
less favorable meteorological conditions, atmospheric emission lines were still apparent in the sky-subtracted
spectra.  Their presence points to sky emission that was significantly variable over the timescale on which the 
telescope was nodded to the reference position.  We corrected for this effect by subtracting the sky spectrum observed 
in the four outermost extraction regions (1, 2, 14, and 15), at the cost of a decreased SNR.}

\noindent The lines detected include 

\noindent (1) A set of 27 \ion{H}{1} recombination lines, consisting of the Pfund $\delta$ and 
Pfund $\epsilon$ lines along with all Humphreys lines 
($n = n_u \rightarrow 6$) with
upper state principal quantum numbers $n_u$ between 15 and 39.

\noindent (2) \gr{10} recombination lines of \ion{He}{2}

\noindent (3) 3 recombination lines of \ion{He}{1}

\noindent (4) The v$=1-0$ $P(1)$ and $P(2)$ transitions of HeH$^+$

\noindent (5) A set of \gr{5} H$_2 $ lines, comprising the v$=1-0$ $O(6)$ and 
and $O(7)$ rovibrational lines; the v$=2-1$ $O(5)$ line, and the $S(13)$ and $S(15)$
pure rotational transitions

\noindent (6) A line at 3.62496~$\mu$m, previously identified by Dinerstein \& Geballe
as a fine structure transition of [\ion{Zn}{4}]

\noindent (7) \gr{5} lines as yet unidentified

\noindent (8) 9 rovibrational lines of CH$^+$: the v$=1-0$ $R(0) - R(3)$ and $P(1) - P(5)$ lines

In Table A1, we list a subset of 52 lines for which the line fluxes could 
be determined reliably, along with the measured fluxes.   {Here, as in Table 4,
the measured values were obtained from direct integration over the line, without any assumption that the line profile is Gaussian.} 

\renewcommand\thetable{A\arabic{table}} 
\setcounter{table}{0}

\begin{deluxetable}{lcr}
\tablecaption{List of detected lines}

\tablehead{
Line & Rest wavelength & Flux\phantom{000} \\
& ($\mu$m) 	   & ($\rm 10^{-18}\,W\,m^{-2}$)}
\startdata
Unidentified                &       3.2817\phantom{0} & $ 14.58\pm  0.51$$^a$\\
\ion{H}{1} 9--5                     &       3.29699 & $332.97\pm  0.81$\\
\ion{H}{1} 38--6                    &       3.36626 & $  3.83\pm  0.24$\\
\ion{H}{1} 37--6                    &       3.37099 & $  6.86\pm  1.69$\\
\ion{H}{1} 36--6                    &       3.37612 & $  8.02\pm  1.00$\\
Unidentified                &       3.3841\phantom{0} & $  5.86\pm  0.55$\\
\ion{H}{1} 34--6                    &       3.38784 & $  4.01\pm  0.62$\\
\ion{H}{1} 32--6                    &       3.40194 & $  6.77\pm  0.23$\\
\ion{H}{1} 31--6                    &       3.41009 & $  7.42\pm  0.40$\\
\ion{H}{1} 30--6                    &       3.41911 & $  8.26\pm  0.30$\\
H$_2$ (2--1)$O(5)$          &       3.43787 & $  1.08\pm  0.43$\\
\ion{H}{1} 27--6                    &       3.45285 & $  9.55\pm  0.22$\\
\ion{H}{1} 26--6                    &       3.46697 & $  8.62\pm  0.53$\\
Unidentified                &       3.4690\phantom{0} & $  1.93\pm  0.41$\\
\ion{H}{1} 25--6                    &       3.48296 & $ 13.68\pm  0.74$\\
\ion{He}{2} 17--10                 &       3.48401 & $ 20.27\pm  1.00$\\
\ion{He}{2} 24--11                 &       3.49012 & $  5.01\pm  0.42$\\
H$_2$ (1--0)$O(6)$          &       3.50081 & Blend \phantom{0}$^b$\\
\ion{H}{1} 24-6                     &       3.50116 & $ 15.62\pm  0.25$$^c$\\
HeH$^+$ (1--0)$P(1)$        &       3.51629 & $  1.55\pm  0.16$\\
\ion{H}{1} 23--6                    &       3.52203 & $ 13.86\pm  0.23$\\
\ion{He}{2} 13--9                  &       3.54432 & $ 53.72\pm  0.21$\\
\ion{H}{1} 22--6                    &       3.54610 & $ 15.44\pm  0.22$\\
CH$^+$ (1--0)$R(2)$         &       3.54958 & $  1.88\pm  0.17$\\
\ion{H}{1} 21--6                    &       3.57410 & Blend \phantom{0}$^b$\\
\ion{He}{2} 23--11                 &       3.57458 & $ 16.25\pm  0.26$$^c$\\
CH$^+$ (1--0)$R(1)$	        &       3.58113 & $  3.68\pm  0.35$\\
\ion{H}{1} 20-6                     &       3.60697 & $ 19.58\pm  0.79$\\
HeH$^+$ (1--0)$P(2)$        &       3.60776 & $  2.08\pm  0.31$ \\
CH$^+$ (1--0)$R(0)$	        &       3.61461 & $  2.34\pm  0.18$\\
$[$\ion{Zn}{4}$]$                  &       3.62496 & $ 43.35\pm  0.39$\\
H$_2$ (0--0)$S(15)$         &       3.6263\phantom{0} & $  2.38\pm  0.33$\\
\ion{H}{1} 19-6                     &       3.64592 & $ 23.86\pm  0.22$\\
\ion{He}{2} 22--11                 &       3.67594 & $  8.34\pm  0.68$\\
CH$^+$ (1--0)$P(1)$         &       3.68757 & $  3.98\pm  0.53$\\
\ion{H}{1} 18-6                      &       3.69263 & $ 24.34\pm  0.42$\\
\ion{He}{1}~$5^3D-4^3P^o$           &       3.70355 & $  8.44\pm  0.48$\\
CH$^+$ (1--0)$P(2)$         &       3.72716 & $  6.06\pm  0.53$\\
\ion{He}{2} 16--10                 &       3.73903 & $ 26.68\pm  1.73$\\
\ion{H}{1} 8--5                     &       3.74056 & $424.41\pm  1.14$\\
\ion{H}{1} 17--6                    &       3.74939 & $ 26.75\pm  0.70$\\
CH$^+$ (1--0)$P(3)$         &       3.76890 & $  7.71\pm  0.52$\\
\ion{He}{2} $21-11$                       &       3.79942 & $  6.66\pm  0.47$\\
H$_2$(1--0)$O(7)$           &       3.80740 & $  5.88\pm  0.47$\\
Unidentified                &       3.8102\phantom{0} & $ 12.85\pm  0.52$\\
CH$^+$ (1--0)$P(4)$         &       3.81286 & $  8.01\pm  0.59$\\
\ion{H}{1} 16--6                    &       3.81945 & $ 34.15\pm  0.54$\\
H$_2$ (0--0)$S(13)$         &       3.8460\phantom{0} & $  1.32\pm  0.80$\\
CH$^+$ (1--0)$P(5)$         &       3.85912 & $ 10.66\pm  0.65$\\
Unidentified                &       3.8675\phantom{0} & $  3.18\pm  0.70$\\
\ion{H}{1} 15--6                    &       3.90755 & Blend \phantom{0}$^b$\\
CH$^+$ (1--0)$P(6)$         &       3.90776 & $ 45.76\pm  0.97$$^c$\\

\enddata
\tablenotetext{a}{$1 \sigma$ statistical errors}
\tablenotetext{b}{Blended with the line
listed immediately below.
The flux entry immediately below lists the total
flux for both transitions}
\tablenotetext{c}{Blended with the line listed
immediately above: entry lists the total
flux for both transitions}

\end{deluxetable}

\renewcommand\thefigure{A\arabic{figure}} 
\setcounter{figure}{0}

\begin{figure}
\includegraphics[scale=0.75,angle=0]{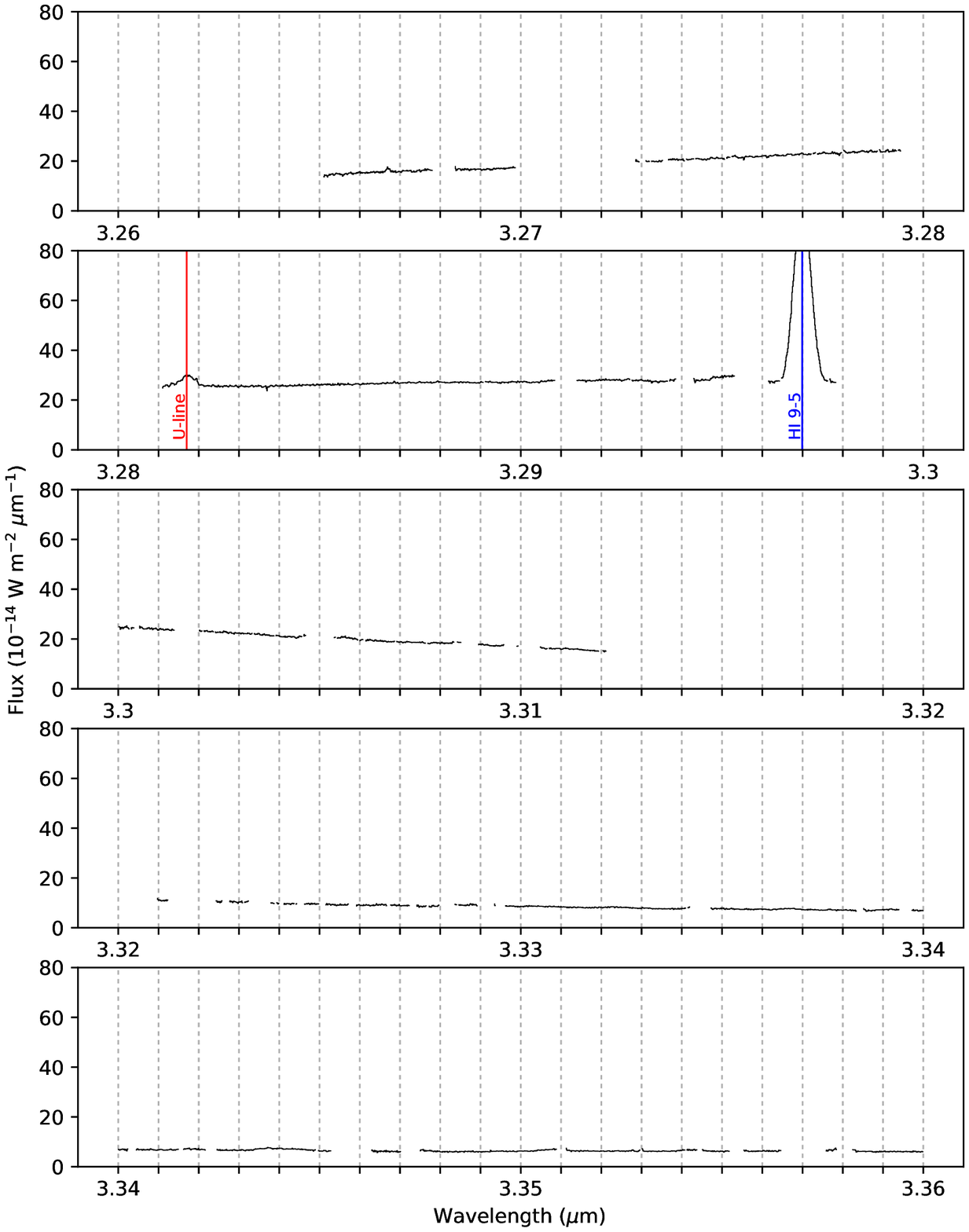}
\caption{NGC 7027 spectra obtained for the 3.26 - 3.36$\,\mu$m spectral region by
summing over the 11 central extraction regions (numbered 3 -- 13 in Figure 1).  The wavelength scale 
is shifted to the rest frame of the source at $v_{\rm LSR} = 25\,\rm km\, s^{-1}$.  Spectral regions that were
severely affected by atmospheric absorption have been excised, resulting in gaps.}
\end{figure}

\begin{figure}
\includegraphics[scale=0.75,angle=0]{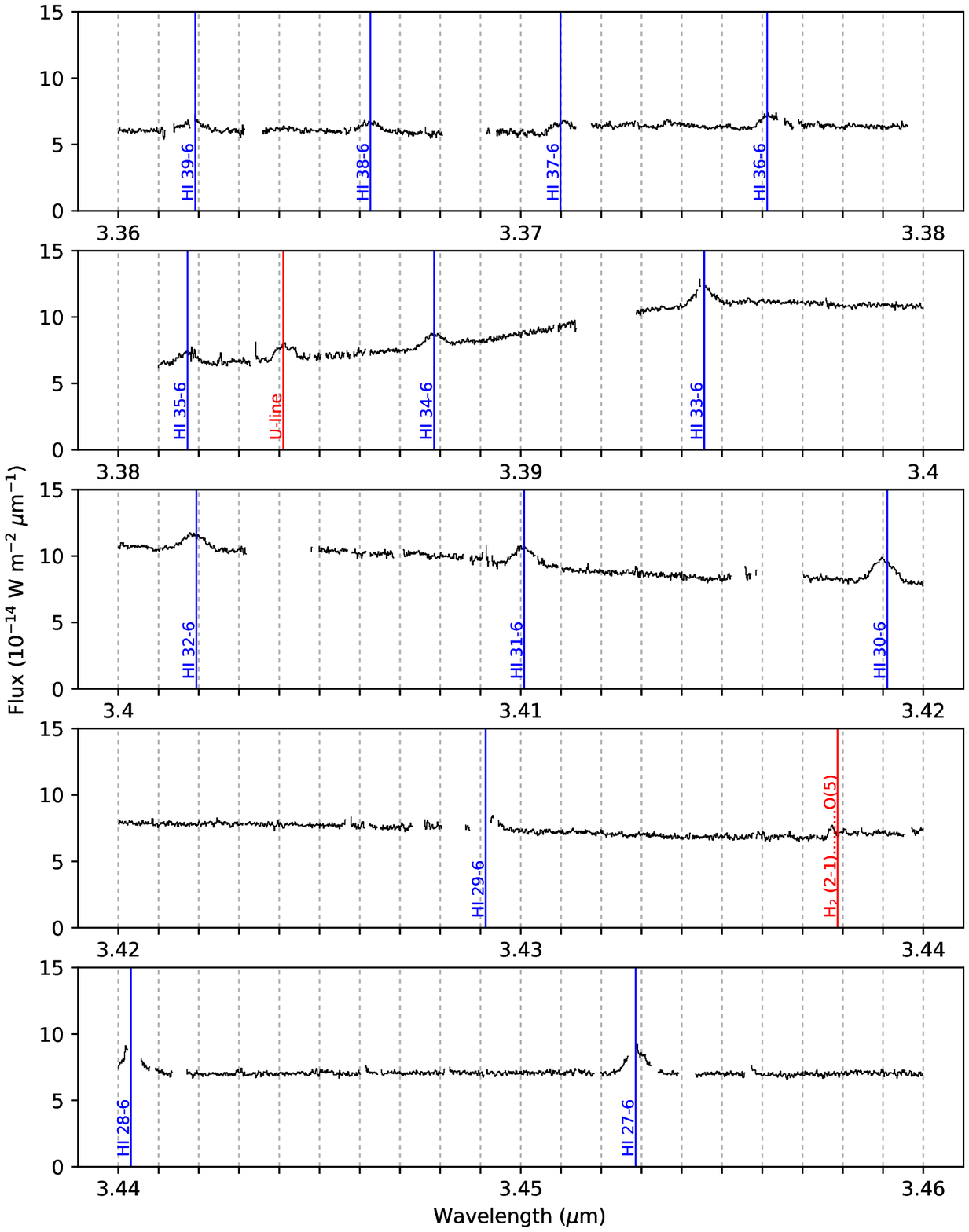}
\caption{Same as Figure A1, but for the 3.36 - 3.46$\,\mu$m spectral region.}
\end{figure}

\begin{figure}
\includegraphics[scale=0.75,angle=0]{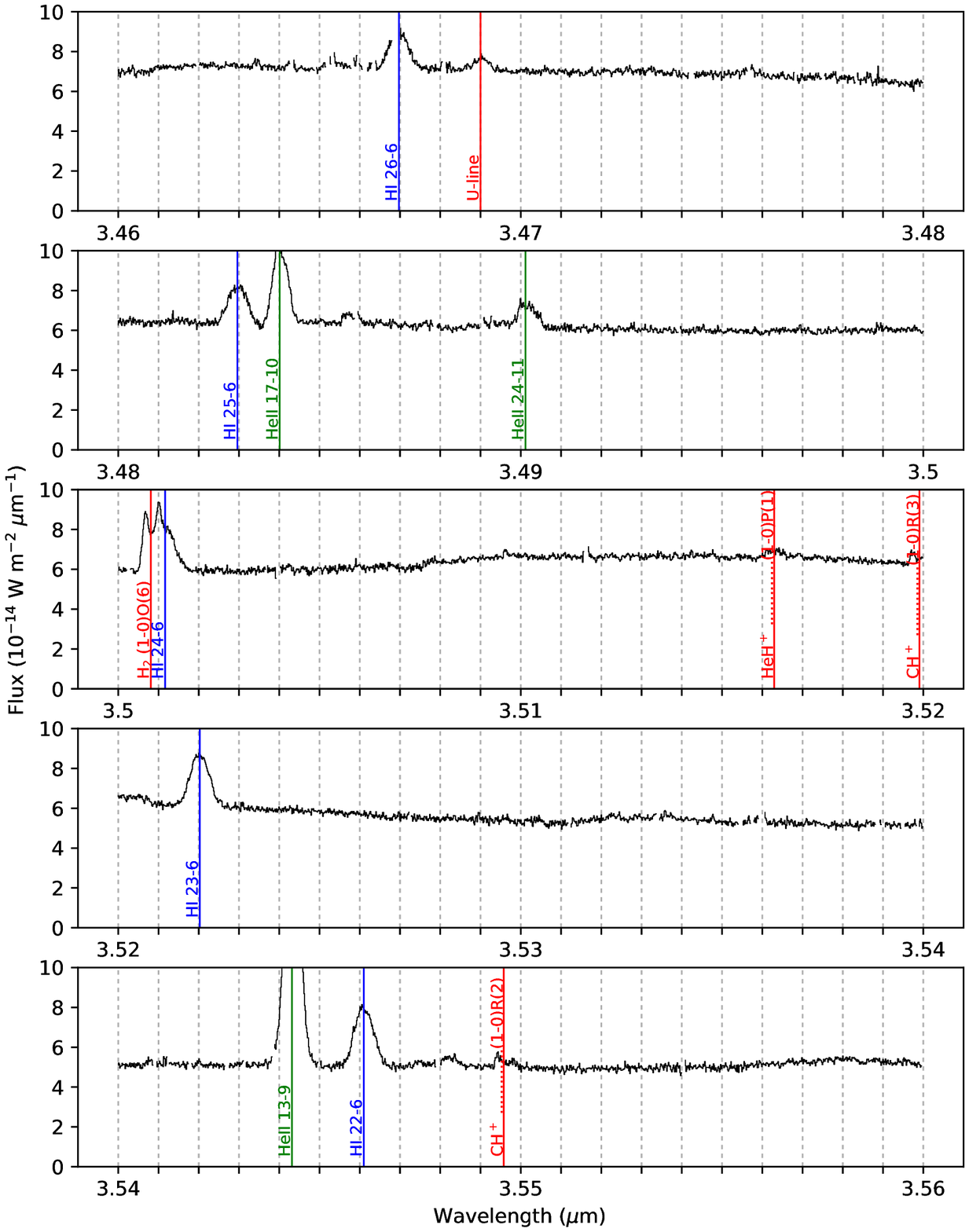}
\caption{Same as Figure A1, but for the 3.46 - 3.56$\,\mu$m spectral region.}
\end{figure}

\begin{figure}
\includegraphics[scale=0.75,angle=0]{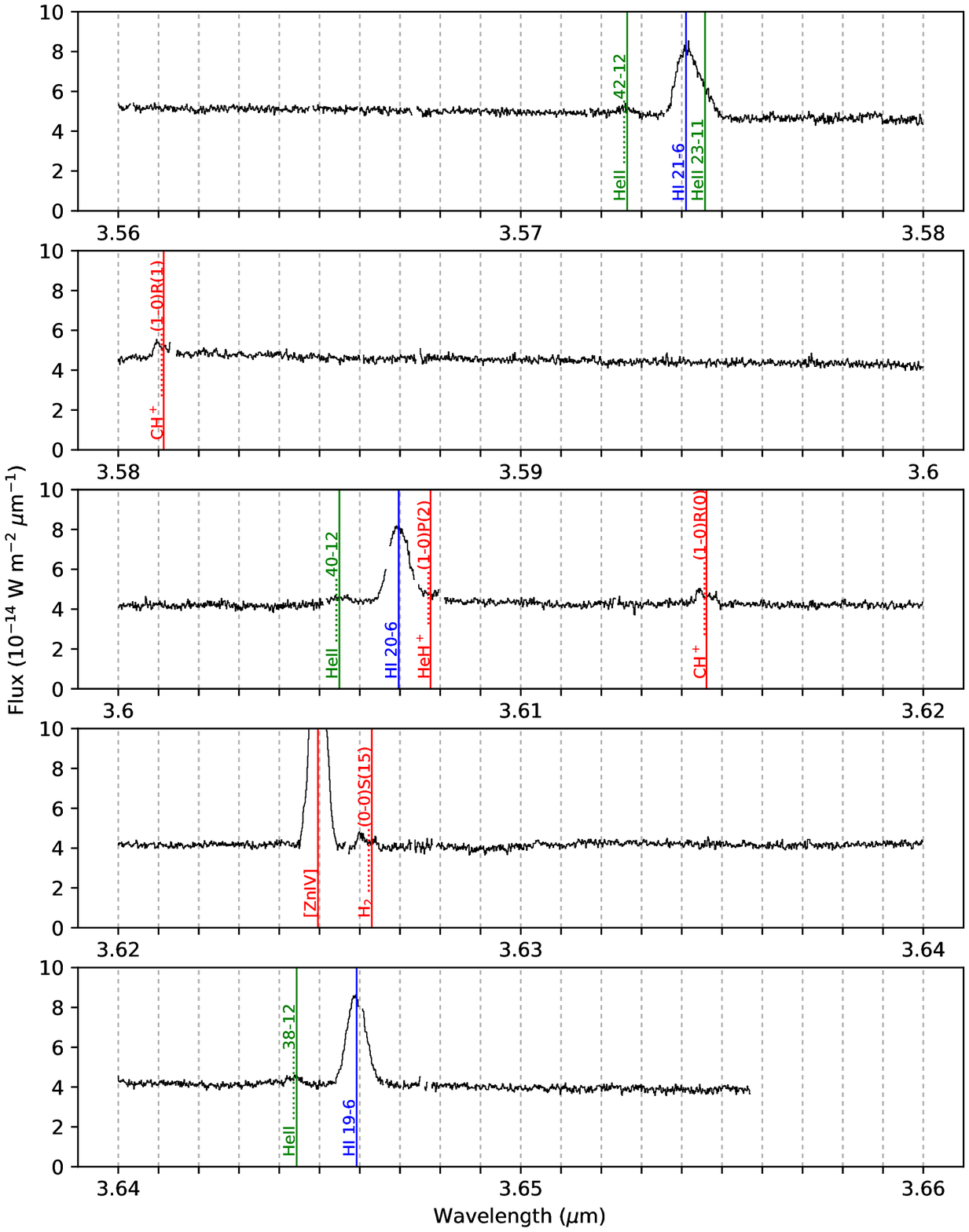}
\caption{Same as Figure A1, but for the 3.56 - 3.66$\,\mu$m spectral region.}
\end{figure}

\begin{figure}
\includegraphics[scale=0.75,angle=0]{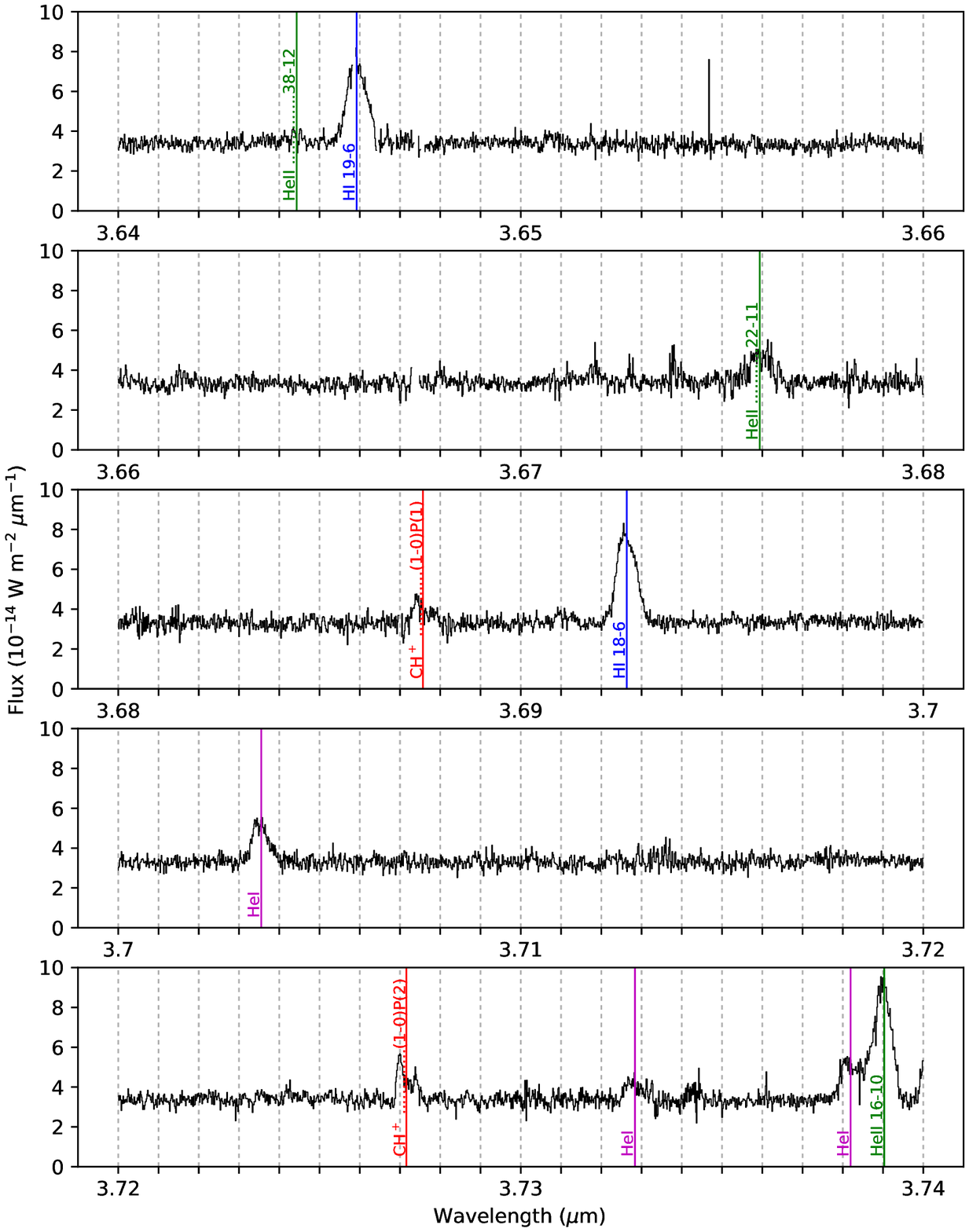}
\caption{Same as Figure A1, but for the 3.64 - 3.74$\,\mu$m spectral region.}
\end{figure}

\begin{figure}
\includegraphics[scale=0.75,angle=0]{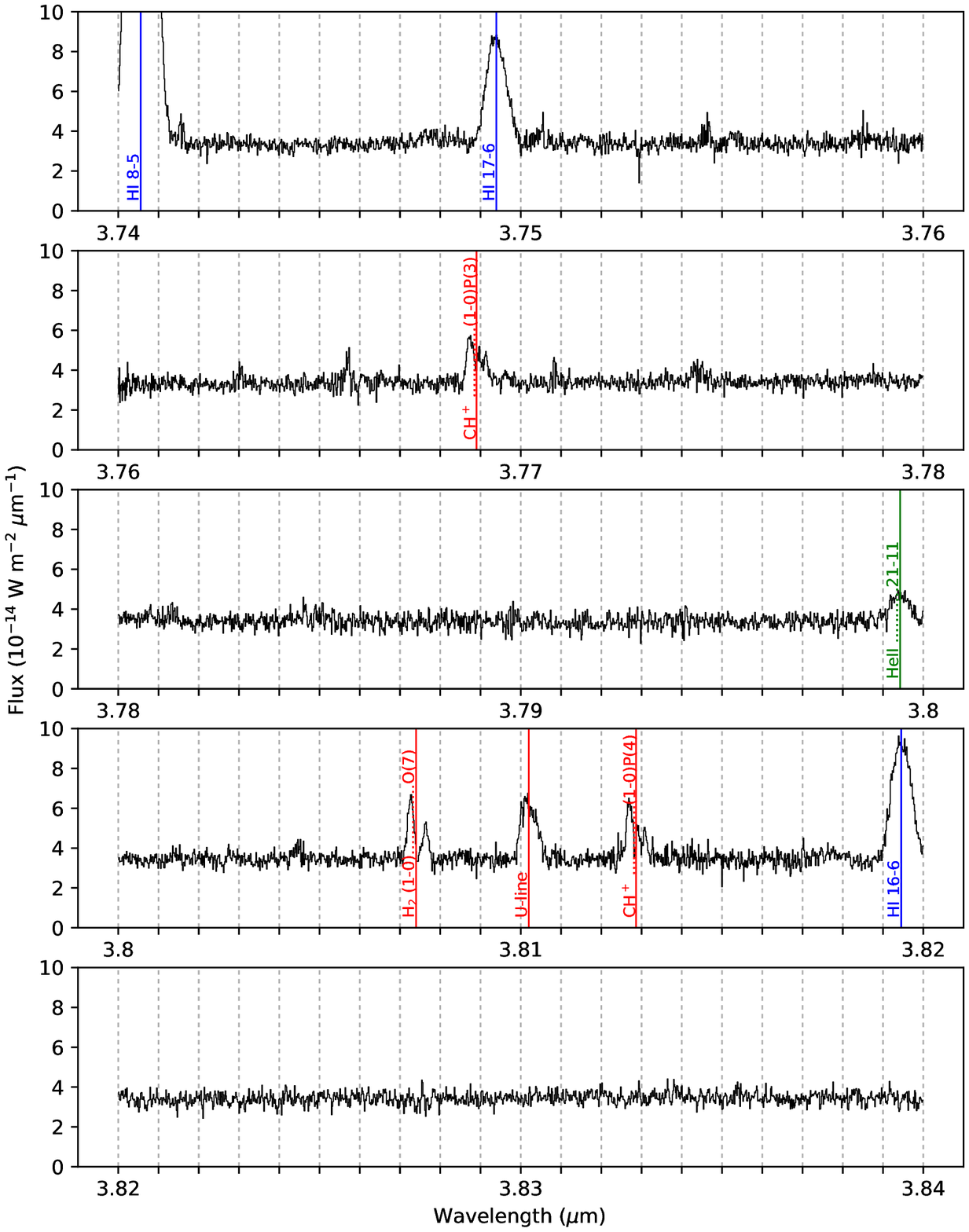}
\caption{Same as Figure A1, but for the 3.74 - 3.84$\,\mu$m spectral region.}
\end{figure}

\begin{figure}
\includegraphics[scale=0.75,angle=0]{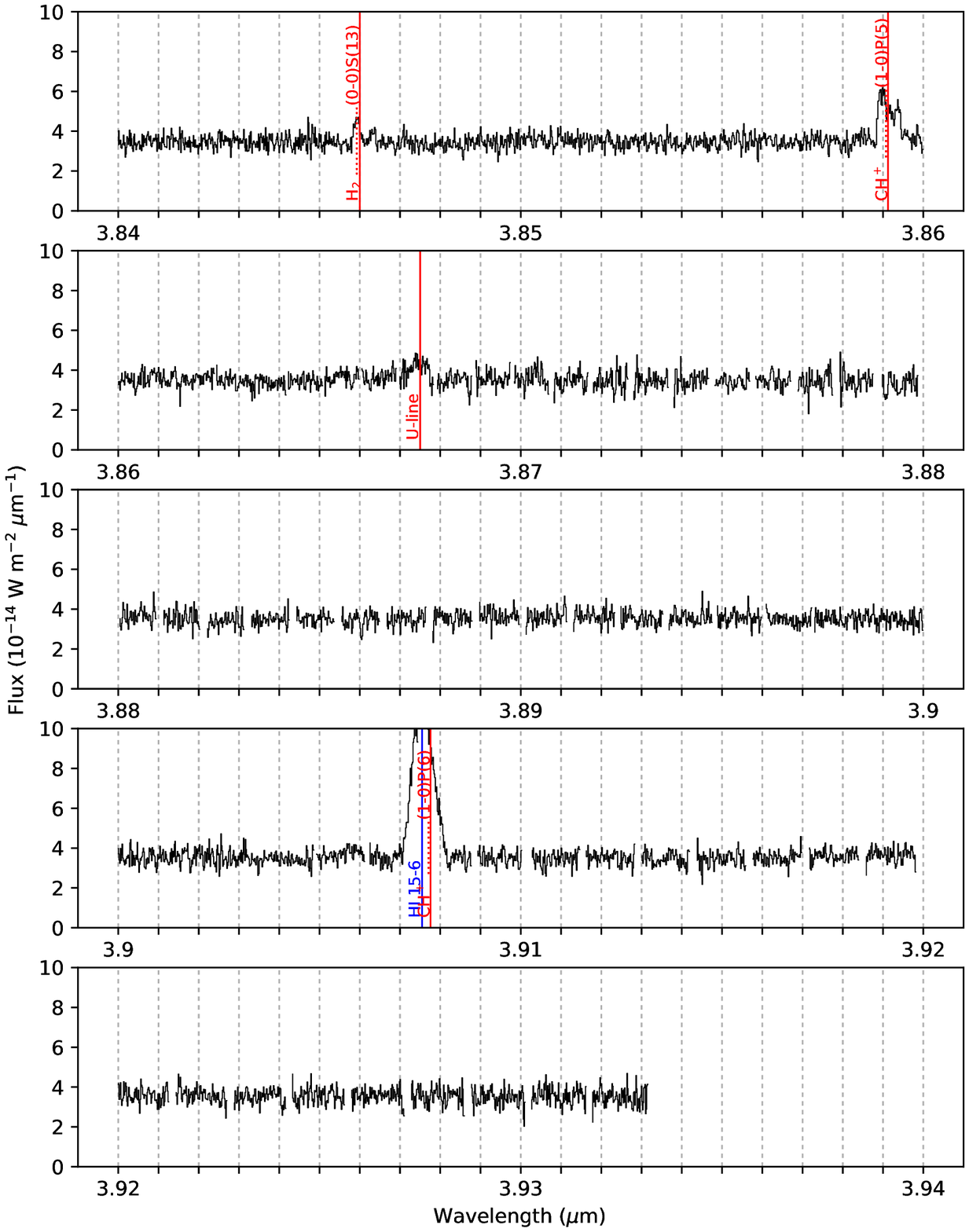}
\caption{Same as Figure A1, but for the 3.84 - 3.94$\,\mu$m spectral region.}
\end{figure}

\end{document}